\documentclass[useAMS,usenatbib]{mnras}

\usepackage{graphicx}
\usepackage{amssymb}
\usepackage{url}
\usepackage{multirow}
\usepackage{times}
\usepackage{txfonts}
\usepackage{aas_macros} 

\usepackage[dvipsnames,usenames]{color}

\title{GeV emission of gamma-ray binary with pulsar scenario}
\author[HU XINGXING et al.]{HU XINGXING$^{1}$\thanks{E-mail:huxx09791@hust.edu.cn},
TAKATA JUMPEI$^{1}$\thanks{E-mail:takata@hust.edu.cn} and TANG QINGWEN$^{2}$\thanks{E-mail:qwtang@ncu.edu.cn} \\
$^{1}$ Institute of particle Physics and astrophysics, School of Physics,
Huazhong University of Science and Technology, Wuhan, Hubei, China\\
$^{2}$ Department of Physics, School of Sciences, Nanchang University, Nanchang 330031, China
}

\begin{document}
\label{firstpage}
\pagerange{\pageref{firstpage}--\pageref{lastpage}}
\maketitle

\begin{abstract}
 We study GeV emission from gamma-ray binaries by assuming that the compact object is a young pulsar. We assume that the relativistic unshocked pulsar wind with Lorentz factor of $10^{4-5}$ can produce the GeV emission by the inverse-Compton scattering process in  the dense soft-photon field of the companion star. The travel distance of the unshocked pulsar wind that moves toward the observer depends on the orbital phase of the pulsar. We discuss that the orbital modulation of the GeV emission is a result of combination of the effects of the travel distance of the unshocked pulsar wind and of the anisotropic soft-photon field of the companion star. In this paper, we study how the effect of the travel distance of the unshocked pulsar wind affects to the orbital modulation of GeV emission.  We apply our scenario to two gamma-ray binaries, LMC~P3 and 4FGL J1405.5-6119. We find that with the suggested system parameters of LMC~P3,  the observed  amplitude of the orbital modulation and the peak width are more consistent with the model light curve by taking into account the effect of the travel distance. For LMC~P3, we analyze the GeV spectrum with 8-years $Fermi$-LAT data and discuss the broadband emission process in X-ray to TeV energy bands. We predict a possible system geometry for 4FGL~J1405.5-6119 by fitting the  GeV light curve.
\end{abstract}


\section{Introduction}

Gamma-ray binary is a  system composed of a massive main-sequence star and a compact object (neutron star or black hole), and   its non-thermal emission has a peak around 1~MeV in a $\nu F_\nu$ of the spectral energy distribution \citep{2013A&ARv..21...64D}. The multi-wavelength observations have confirmed about 10~ gamma-ray binary systems, namely, PSR B1259-63/LS~2883 \citep{1992ApJ...387L..37J,1994surp.rept.....C,2005A&A...442....1A},
PSR J2032+4127/MT91~213 \citep{2015MNRAS.451..581L,2017MNRAS.464.1211H,2017ApJ...836..241T,2018ApJ...867L..19A}, LS~5039 \citep{2012A&A...548A.103M,2009ApJ...697..592T,2006A&A...460..743A,2009ApJ...706L..56A},
LS~I+61$^{\circ}$303 \citep{2006Sci...312.1771A,2013ApJ...779...88A},1FGL~J1018.6-5856 \citep{2012A&A...541A...5H,2015ApJ...806..166A}, H.E.S.S. J0632+057 \citep{2009ApJ...690L.101H,2017ApJ...846..169L,2018PASJ...70...61M},
LMC~P3 \citep{2016ApJ...829..105C} ,4FGL~J1405.4-6119 \citep{corbet19} and HESS J1832-093 \citep{2020arXiv200102701M,2020arXiv200107138T}. The gamma-ray binary is thought to be a short phase before high-mass X-ray binary in the binary evolution.

The gamma-ray binaries are  probably divided into two groups based on the nature of the companion star, Be/Oe-type (PSR B1259-63/LS~2883, PSR J2032+4127/MT91~213, LS~I+61$^{\circ}$303 and H.E.S.S. J0632+057)
or O-type (LS~5039, 1FGL~J1018.6-5856, LMC P3 and 4FGL~J1405.5-6119).
The periods of gamma-ray binaries with the Be-type companion star ($P_{orb}\sim$ several years, except for $P_{orb}\sim 30$ days
of LS~I+61$^{\circ}$303) is longer than that  with the O-type companion star  ($P_{orb}\sim$ several days  to about ten days).  The orbit of the compact object moving around  the Be-type companion star is  elongated with an eccentricity  $e>0.8$  (except for $e\sim 0.6$
of LS~I+61$^{\circ}$303), while the eccentricity  of the compact object circulating  around  the O-type companion star is $e<0.5$. One interesting property of the  emission from the gamma-ray binaries is  that its observed intensity varies along the orbit.
For the system with a Be-type companion star, the compact object will interact with the Be-disk twice in one orbit,
and it has been observed that the  emissions in X-ray and possibly in TeV bands are  enhanced during the interaction \citep{2009MNRAS.397.2123C,2015MNRAS.454.1358C,2019A&A...627A..87C}. A flare-like  GeV emission from PSR B1259-63/LS~288 has
been detected after the second interaction between the pulsar and Be-disk \citep{abdo11,tam11,tam18}.
For the binary system with an O-type compact star, the  X-ray/GeV and TeV bands
would show an enhancement around the superior conjunction (SUPC) or inferior conjunction (INFC) of the orbit of the compact object and the GeV emission is
observed over the entire orbit \citep{2016MNRAS.463..495C,2015ICRC...34..885M,2020arXiv200102701M}  .

Because of the recent discoveries of new  gamma-ray binaries hosting O-type companion star (LMC~LP3 and 4FG J1405.4-6119),
the GeV emission properties of the binary systems  have been revealed.  The  profile of LS~5039, for example,  has  a peak at the SUPC
of the compact star's orbit \citep{2009ApJ...706L..56A,2016MNRAS.463..495C}, while it of LMC~P3 is shifted
from the SUPC  toward the periastron  \citep{2016ApJ...829..105C,2019MNRAS.484.4347V} ; no orbital parameter except for the period is available for 4FG J1405.4-6119 \citep{corbet19}.
The light curve of LS~5039 is described by  a single broad peak,
while LMC~P3 will show an asymmetric pulse shape (see Figure~\ref{fig:lcumev2}),
and 4FGL J1405.4-6119 probably shows a double peak structure (Figure~\ref{fig:J1405}). These properties of the light curve  will
provide an additional information of the GeV emission from the gamma-ray binaries hosting O-type companion star.
Except for 1FGL J1018.6-5856, moreover, the peak position of the GeV emission of other three gamma-ray binaries is shifted from the peaks
of the X-ray/TeV bands~\citep{2016MNRAS.463..495C, 2016ApJ...829..105C,corbet19} (no TeV observations for 4FGL J1405.4-6119 have been reported). This  indicates
that the GeV emission process are different from the processes of X-ray emission and TeV emission.

In this study, we discuss the model  that the compact object of the gamma-ray binary  is a  young pulsar. In the pulsar scenario, the pulsar
wind, which is composed of the electron/positron and magnetic field, interacts with the stellar wind/stellar disk.
The X-ray and TeV emissions are produced by the synchrotron and inverse-Compton scattering process, respectively, of the
relativistic pulsar wind particles accelerated at the termination shock \citep{1997ApJ...477..439T} . The orbital modulation of
the emission in the X-ray bands  attributes to the Doppler boosting effect due to the shocked pulsar
wind~\citep{2010A&A...516A..18D,2014ApJ...790...18T}, and it in TeV bands is caused by the Doppler boosting effect plus
the effects of the anisotropic photon field and of the pair-creation process.

The origin of the GeV emission from the gamma-ray binaries remains to be solved. It has been established that young pulsars emit the GeV gamma-rays in the
magnetosphere \citep{2009ApJ...706L..56A}. However, the GeV emission
modulating along the orbital phase of the gamma-ray binaries
will not be explained by  the magnetospheric  emission. Within the framework of the pulsar binary system, the GeV emission modulating with the orbital phase
would  be produced by the Doppler boosting  of the  synchrotron photons due to the finite speed of the shocked pulsar wind \citep{2017ApJ...838..145A} or the inverse-Compton scattering  (hereafter ICS) of the cold relativistic pulsar wind \citep{2008AIPC.1085..253S,2010tsra.confE.193K,2011ASSP...21..531T,2014ApJ...790...18T} or
ICS the shocked pulsar wind \citep{2012ApJ...761..146Y,2013A&A...551A..17Z} off the soft-photons from the companion star.

In this paper, we assume that the GeV emission of the gamma-ray binaries is  produced
by the ICS process between the cold-relativistic pulsar wind (hereafter unshocked pulsar wind) and the soft photon from the companion star.
The orbital modulation of the ICS of the  pulsar wind, especially  for LS~5039
 system, are discussed in the previous studies \citep{ball00,2013A&A...551A..17Z,2014ApJ...790...18T}.
 Since the soft photon field is anisotropic in the emission region and depends on the distance from the companion star to the emission region, the emissivity of
 ICS varies along the orbital phase.
 For the circular orbit, for example, the ICS intensity tends to be  the
 maximum value at the SUPC  and  the minimum at the INFC, since the scattering  processes at SUPC and INFC
 are caused by a head-on collision process and a tail-on collision process, respectively.
 For an elongated orbit, since the soft-photon density on the orbit of the pulsar is
 maximum at the periastron, the orbital position of the intensity maxima  would shift toward  the periastron. In the previous
 studies, however, the predicted light curve has a single peak in the light curve and does not discuss  the formation of the double peaks.
 An  additional effect, therefore, will be required to  explain the orbital variation of  the GeV emissions of
 LMC~LP3 and especially for 4FG J1405.4-611.

 In addition to the effect of the anisotropic soft-photon field, we will explore the orbital modulation caused by the dependency of travel distance
 of the unshocked pulsar wind that moves toward the observer on the orbital phase.
 This effect is introduced in \cite{2008MNRAS.385.2279S} for PSR~B1259-63/SS2883 system and in  \cite{2008AIPC.1085..253S,2008MNRAS.383..467K} for the very high-energy emission (TeV emission)
 of LS~5039.  Since the interaction between the  pulsar wind and stellar wind/disk will produce a termination-shock with cone-like
 structure \citep{1996ApJ...469..729C}, the distance from the pulsar to the termination-shock  ($r_s$ in Figure~\ref{fig:shape}) increases  with the angle measured from the axis created by the pulsar  and the companion star. This causes a dependency of the distance from the pulsar to the termination shock in the direction of the  observer on the orbital phase and results in the orbital dependent intensity of the ICS of the unshocked pulsar wind. For the gamma-ray binary with O-type star, the GeV light curve is not well studied with
 the effect of the shock geometry and the system parameters. In this paper, therefore, we will perform an  investigation   for  the modulation of
 the GeV emission with the shock-cone model, and will apply the model to two  gamma-ray binaries,
 LMC~P3 and 4FGL~J1405.4-6119.

 LMC~P3 is the first gamma-ray binary  outside Milky~Way and it is discovered in the Large Magellanic Cloud (LMC)  \citep{2016ApJ...829..105C}, which   is a neighbor galaxy located at about 50~kpc from the Earth \citep{2006ApJ...652.1133M,2013EAS....64..305P,2014AJ....147..122D}. The companion star in the LMC~P3  system is O-type star, but the compact object has not been identified yet. In the LMC, $\sim50$  supernova remnants (SNRs) are  known to emit X-rays \citep{2012ApJ...759..123S}, and LMC~P3 is associated with SNR 0535-67.5 in the H~II region DEM L241~\citep{1976MmRAS..81...89D}. \cite{1981ApJ...248..925L} reports the first detection of the X-ray emission from the SNR~0535-67.5.  \cite{2012ApJ...759..123S} find a compact X-ray source (CXOU~J053600.0-673507) with a luminosity of $\sim 2-3\times 10^{35}{\rm erg~s^{-1}}$ in 0.3-10keV energy bands and an optical counterpart,
 O5III(f) star. They therefore suggest that  CXOU~J053600.0-673507 is the high-mass X-ray binary hosting neutron star or black hole. \cite{2016ApJ...829..105C} analyze  the GeV counter part (LMC~P3) of CXOU~J053600.0-673507 in data of $Fermi$ Large Area Telescope ($Fermi$-LAT) and find  a 10.3~day periodic modulation that
reveals LMC~P3/ CXOU~J053600.0-673507 is the binary system with a orbital
period of $P_{orb}\sim 10.3$~day. They also confirm the orbital
modulation of the X-ray emission. \cite{2018A&A...610L..17H} subsequently detect the TeV emission from LMC~P3. \cite{2019MNRAS.484.4347V} find that the binary system is slightly eccentric with an eccentricity $e=0.40\pm 0.07$ and that the mass function, $f=0.0010\pm 0.0004M_{\odot}$, favors a neutron
star as a compact object. These multi-wavelength observations will suggest that  LMC~P3/CXOU~J053600.0-673507 is the gamma-ray binary system hosting an energetic
young pulsar. The orbital parameters determined by \cite{2019MNRAS.484.4347V}  suggest that the GeV emission
has a peak intensity at a phase between  the superior conjunction and periastron, while TeV emission (and probably X-ray emission) shows the maximum intensity at around the
inferior conjunction.

4FGL~J1405.1-6119 is the recently discovered gamma-ray binary with an  orbital period of $P_{orb}\sim 13.7135$ days\citep{corbet19}  and is hosting O6.5~III companion star. The
estimated distance with using tabulated absolute magnitude for an O6.5~III star is $d\sim 6.4-8.9$kpc \citep{corbet19} . The GeV light curve is composed of a strong narrow peak plus
a small peak. \cite{corbet19}  reveal that the X-ray emission (CXOGSG J140514.4-611827) also modulates
with the orbital phase.  The position of the main peak of the GeV emission
is shifted from  the X-ray peak that would  be described by a single broad peak.
In this paper, since the orbital parameters have not been reported,
we predict a  possible geometry of the binary system   by fitting GeV light curve with the model.

In this paper, we will discuss the multi-wavelength emission properties.  \cite{2016A&A...586A..71A} report the  GeV spectrum of LMC~P3 with six years of $Fermi$-LAT observations. We also redo  the spectral analysis of LMC~P3 with the eight years of the $Fermi$-LAT observations, since the spectral information in the GeV bands is important to understand the multi-wavelength emission process from X-ray to TeV energy bands.  In section~\ref{fermi}, we extract the
GeV spectrum of LMC~P3 using $Fermi$-LAT data. In section~\ref{model}, we will describe our theoretical model for the GeV emissions from the unshocked pulsar wind and X-ray/TeV emissions from the shocked pulsar wind.  We discuss the dependency of the GeV light curve
with the system parameters and compare the results with the observations for LMC~P3 and 4FGL~J1405.1-6119 in section~\ref{obser}. A
brief summary will be provided  in section~\ref{summary}.

\section{Fermi data analysis}
\label{fermi}
\subsection{Event Selections and Background Subtraction}
Gamma-ray events for LMC~(RA.=80.894, Decl.=-69.756) are derived from the {\it Fermi} Science Support Center\footnote{https://fermi.gsfc.nasa.gov/ssc/}, which span the period from
4th Aug 2008 to 4th Aug 2016. Due to low background contamination, we cut the low-energy events,
such as below 200~MeV, and select them up to 100~GeV, which lead to a 10$^\circ\times$10$^\circ$ box region~(ROI) being suited in this gamma-ray band. Events with zenith angles $>$90$^\circ$ are excluded to eliminate the contamination from the Earth-limb gamma rays. Instrument response function (IRSF) of P8R2$\_$SOURCE$\_$V6 is employed in our data analysis.

As discussed in \cite{2017ApJ...843...42T} and \cite{2018Ap&SS.363...25T}, the G template, which includes four gaussian-disk sources and four point sources, is a good template for recovering the gamma rays observed by Fermi-LAT in LMC field. In the G template, LMC P3 is near the centre of the G template~(RA.=80.894, Decl.=-69.756). For source of LMC P3, other 7 sources of the G template~(G1, G2, G3, G4, P1, P2 and P4) and 3 point sources~(3FGL J0601-7036, 3FGL J0529-7242 and 3FGL J0437-7330) are the background sources with spectral parameters free, while 3FGL sources, that are out of ROI but 15$^\circ$ around LMC centre, are the buffering background sources with all spectral parameters fixed. In addition, a Galactic diffuse source and an isotropic gamma-ray source are added, which are represented by ``gll$\_$iem$\_$v06.fits'' and ``iso$\_$P8R2$\_$SOURCE$\_$V6$\_$v06.txt'' respectively.
\subsection{Spectral analysis for LMC P3}
Gamma-rays between 200~MeV and 100~GeV are divided into 11 logarithmic energy bins. A single power-law spectrum (PL) is assumed for LMC~P3, while all background sources are modeled by corresponding spectral functions in \citet{2016A&A...586A..71A} \& \citet{2017ApJ...843...42T},   i.e., PL for 5 LMC sources (G2, G3, G4, P2 and P4), power law with exponential cutoff (PLC) for the P1, log-parabola (LP) for the G1.  A global fit is performed in a python package (Fermipy)\footnote{https://fermipy.readthedocs.io/en/stable}\citep{2017ICRC...35..824W}, whose resultant parameters are fixed in the afterward spectral data fits except for the normalization of the LMC~P3. Performing the binned maximum-likelihood analysis, we report the flux value and the upper-limit flux (95\% C.L.), which are plotted in Fig.~\ref{fig:spc}. Our results are consistent with that reported in \citet{2016A&A...586A..71A}.

\section{Theoretical Model}
\label{model}
\begin{figure}
	\includegraphics[scale=0.6]{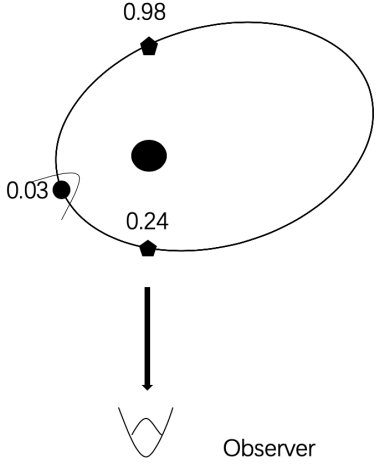}
	\caption{Schematic view for LMC~P3. The pulsar is orbiting around the companion star (large filled circle) with an eccentricity  of  $e=0.4$.  An interaction of the pulsar wind and stellar wind creates a shock, which wraps the pulsar. In section~4.2, we assume  that the orbital phases $\phi$=0.03, 0.24 and 0.98 corresponds to the periastron, INFC and SUPC, respectively, of the pulsar orbits, where the phase zero refers the time  MJD~57,410.25. \citealt{2019MNRAS.484.4347V}. This orbital parameters are chosen  to explain the observed GeV modulation with the inverse-Compton process of the cold-relativistic pulsar wind. }
	\label{fig:obt}
\end{figure}

In our theoretical  model, we assume that the compact object is a pulsar, which produces strong pulsar wind interacting with the stellar
wind (Figure\ref{fig:obt}). The unshocked pulsar wind scatters off the soft photons from the companion star and produces GeV emission that modulates along
the orbital phase. The interaction between the relativistic pulsar wind and the stellar wind creates the termination shock. At the termination shock, the pulsar wind particles (electrons and positrons) are accelerated to higher energy
and produce the X-ray and TeV gamma-ray via  the synchrotron and the ICS  process, respectively.

\subsection{Emission from unshocked  pulsar wind}
We assume that (i)  the unshocked pulsar wind is isotropic,  (ii) it is composed of the electron/positron and magnetic field and
(iii) it carries the spin-down power ($L_{sd}$) of the pulsar.  To describe the particle energy of the pulsar wind, we introduce so called magnetization
parameter that is the ratio of the magnetic energy to the particle energy of the relativistic wind,
\begin{equation}
  \sigma(r) = \frac{B_W^2(r)}{4\pi\Gamma_W(r)u_W(r) n_W(r) m_e c^2},
\end{equation}
where $B_W$ is the magnetic field in the wind region before the shock, $n_W$ is the {\it proper} number density of the electrons/positrons,
$u_W$ and $\Gamma_W =\sqrt{1 +u_W^2}$ are the dimensionless radial four velocity and the Lorentz factor of the unshocked flow, respectively,

Using the conservations of the particles and of the energy, we may relate the Lorentz factor of the relativist flow ($u_W\sim \Gamma_W$) with
the magnetization parameter as described in \citep{1984ApJ...283..694K}

\begin{eqnarray}
  \Gamma_W(r)&=&\frac{L_{sd}}{4\pi n_W(R_{lc})  m_e c^3 R_{lc}^2(1+\sigma)}\nonumber \\
    &\sim &3\times 10^{4}(1+\sigma)^{-1}\left(\frac{\kappa}{10^5}\right)^{-1}\left(\frac{L_{sd}}{10^{37}{\rm erg~s^{-1}}}\right)^{1/2},
\label{gammaw}
\end{eqnarray}
where $R_{lc}$ is the light cylinder radius of the pulsar and  $\Gamma_W n_W(R_{lc})=\kappa N_{GJ}(R_{lc})$ with $N_{GJ}(R_{lc})=B(R_{lc})/(2\pi R_{lc}e)$
being the Goldreich-Julian number density measured at the light cylinder and $\kappa$ the multiplicity. The multiplicity is the number of new positron/electron pairs that a primary particle can make via pair-creation cascade . The multiplicity of the young pulsars will be determined by the pair-creation cascade above the polar cap acceleration and it will not exceed $\sim $a few $\times 10^5$ \citep{2015ApJ...810..144T,2019ApJ...871...12T,2001ApJ...554..624H,2001ApJ...560..871H}.
We expect therefore that the typical Lorentz factor of the unshocked pulsar wind will be of the order of $10^{4-5}$. In this paper,
we assume that the energy conversion from the magnetic energy to the particle energy is almost completed
at vicinity of the pulsar and produces an kinetic energy dominated flow, $\sigma \ll 1$.

In this study, we assume that  the particles in the
unshocked pulsar wind  have a mono-energetic distribution. If the pair-creation
cascade is developed in the pulsar wind, the energy distribution will  be
  modified \citep{2008AIPC.1085..253S}. In the current study, however,
the unshocked pulsar wind has a Lorentz factor  of the order of
$10^{4-5}$ and produce GeV photons via the ICS  process. Since
the optical depth of GeV photons is much less than unity, the pair-creation cascade will not develop in the pulsar wind region.

The radiation  power per unit energy per unit solid angle of single particle is
calculated from
\begin{equation}
  \frac{dP_{IC}}{d\Omega}=\mathcal{D}^2\int \int _0 ^{\theta_c} (1-\beta_W \cos \theta_0) I_b/h  \frac{d\sigma {'}}{d\Omega ^{'}}\cos\theta d\Omega_0,
  \label{ics}
\end{equation}

\begin{equation}
I_{b}=\frac{2\epsilon ^3/h^2c^2}{\mathrm{exp}(\epsilon/kT)-1}
\end{equation}
where $\mathcal{D}=\Gamma_W^{-1}(1-\beta_W\cos\theta_s)$, $\beta_W$ is the wind speed in units of the speed of light, $\theta_0$ (or $\theta_s$) describes the angle between the direction of the particle motion and the propagating direction of the incident (or scattered)  photon and  $\theta_c$ is the angular size
of the companion star measured from the emission region.  In addition,  $h$ is the Planck constant,
$I_b$ is the distribution of background photons, and $ d\sigma {'}/d\Omega ^{'}$ represents
the differential Klein-Nishina cross section.

\begin{figure}
  \includegraphics[scale=0.45]{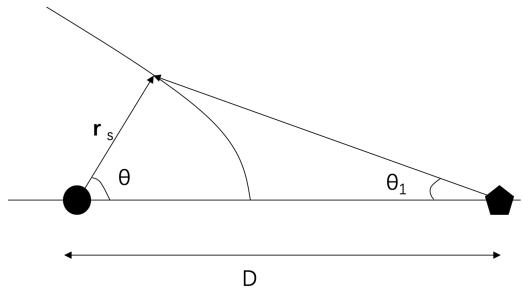}
  \caption{Geometry of the shock cone. The black filled circles and regular pentagon represent the pulsar and the companion star,
respectively. The distance from the pulsar to the shock, $r_s$,  is described by the equation~(\ref{eqv:rsdd}).}
  \label{fig:shape}

\end{figure}

Total power radiated toward the observer by the unshocked pulsar wind
is calculated from
\begin{equation}
P_{IC}=\int \int \frac{dP_{IC}}{d\Omega} \Gamma_W(r) n_W(r)r^2 d\Omega dr,
\end{equation}
where the integration for the distance is taken from the vicinity of the pulsar to the shock in  the direction of the observer.
The modulation of the emission with the orbital phase  is produced by (i) anisotropy of the soft photon at the emission region
and (ii) the dependence of the distance from the pulsar to the termination shock in the direction of the observer on the orbital phase \citep{2007ApJ...671L.145S}. To calculate the travel distance of the unshocked pulsar wind, we explore the shock cone geometry discussed  in \cite{1996ApJ...469..729C} (Figure \ref{fig:shape}),
\begin{equation}
  r_{s}(\theta)=D\sin\theta\textrm{csc}(\theta+\theta_1)
 \label{eqv:rsdd}
\end{equation}
and
\begin{equation}
  \theta_1\textrm{cot}\theta_1=1+\eta(\theta\textrm{cot}\theta-1),
\label{eqv:theta}
\end{equation}
where $D=\frac{0.35 {\rm AU}(1-e^2)}{1+e \cos\phi}$ is the separation between two stars, $\theta$ and $\theta_1$  describe the angle to a position on the  shock measured from the pulsar and companion star, respectively (c.f. Figure~\ref{fig:shape}). In addition, $\eta$ is the ratio of the momenta of the two winds, 
\begin{equation}
  \eta\equiv \frac{L_{sd}}{\dot{M}v_wc},
  \end{equation}
where $\dot{M}$ is the mass loss rate of the outflow from the companion star and $v_w$ is the speed of the outflow.
\cite{1996A&A...305..171P} study the mass loss rate of O-type star
 by fitting the $H\alpha$ profile.  They  find that the mass loss rate
 can be  larger than $10^{-6}M_{\odot}\mathrm{year^{-1}}$, and the
 terminal wind velocity is around $2000$km/s, for  which  the momentum ratio  $\eta$ can be in the order of 0.01 for  a spin down pulsar of the pulsar $L_{sd}\sim 10^{36}{\rm erg~s^{-1}}$.

\begin{figure}
  \includegraphics[scale=0.5]{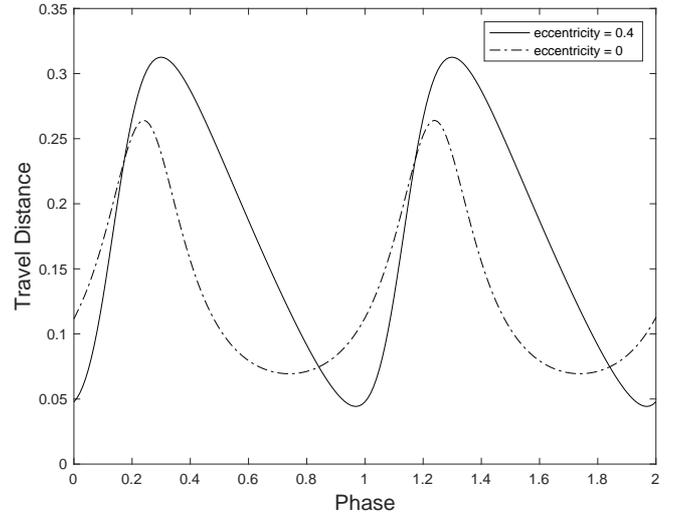}
  \caption{Variation of the travel distance of the unshocked pulsar wind before the shock.
    The geometry of the shock cone is modeled by \citealt{1996ApJ...469..729C}.
    The results are for $\eta =0.03$ and inclination
 angle is $\alpha=30^{\circ}$. The dotted line and solid line are for
the eccentricity $e=0$ and $e=0.4$, respectively. The periastron and INFC is located at
$\Phi_{peri}=0$ and $\Phi_{INFC}=0.24$, respectively.  The SUPC is located
  at $\Phi_{SUPC}=0.76$ for $e=0$ and $\Phi_{SUPC}=0.98$ for  $e=0.4$, respectively.}
  \label{fig:intl}
\end{figure}

\begin{figure}
  \includegraphics[scale=0.5]{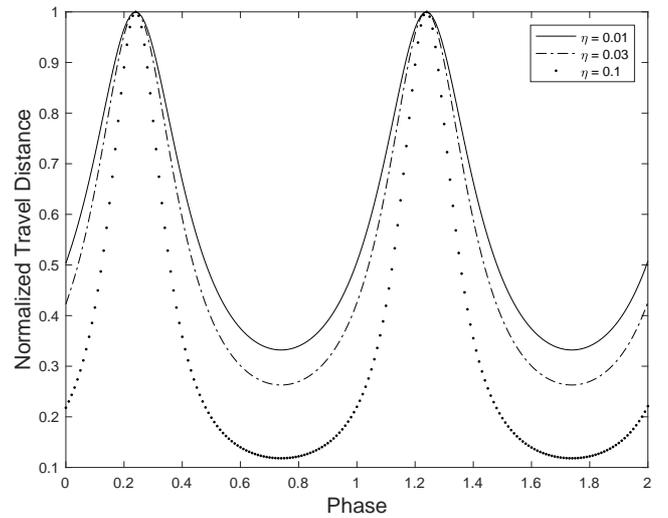}
  \caption{Variation of the travel distance of unshocked pulsar wind with using
    the circular  orbit $e=0$ and the inclination angle $\alpha=30^{\circ}$. The travel distance is normalized by the value at the INFC. The results are for the moment ratio $\eta$=0.01 (solid line),  0.03 (dash-dotted line) and $0.1$ (dotted line), respectively. The physical distance at the INFC is $\sim 0.13$AU for $\eta=0.01$, $\sim 0.26$AU for
$\eta=0.03$ and $\sim 1.0$AU for $\eta\sim0.1$.}
        \label{fig:intl1}
\end{figure}

 Figure~\ref{fig:intl} and \ref{fig:intl1} summarize  the  travel distance of
 the unshocked pulsar wind toward the observer as a function of the orbital phase for the different
 system parameters. In~Figure\ref{fig:intl}, we present the results with
$\eta=0.03$ and system inclination angle $\alpha=30^{\circ}$; the dotted line  and  solid
line are results for the eccentricity $e=0$ and $e=0.4$, respectively: For circular orbit (dashed line), since the separation is constant, no periastron is defined, while  for eccentricity of 0.4 (solid line), we set the periastron to be phase 0. In the figure,
we choose the  phase zero ($\Phi_{orb}=0$)  as the periastron for elongated orbit and
$\Phi_{orb}=0.24$ as  the INFC. With $\Phi_{orb}=0.24$ of the INFC,
 the superior conjunction  is located at $\Phi_{orb}=0.76$ for the orbit with  $e=0$ and
 $\Phi_{orb}\sim 0.98$ for $e=0.4$, respectively. The unshocked pulsar wind travels toward the observer is stopped if the line of sight is outside
 the shock cone. In such a case, the travel distance of the unshocked pulsar wind moving toward the line of sight has a finite value for whole orbital phase,
 as  Figure~\ref{fig:intl} shows.  As we expect for the case of $\alpha>90^{\circ}-\theta_{shock}$, the line of sight at around the INFC is within
 the shock cone and   the pulsar wind moving  toward the observer will not be stopped  by the shock. We note that if there is a back shock proposed by
\cite{2012A&A...544A..59B}, the pulsar wind in all direction would be stopped by the vicinity of the pulsar.
 In this paper, however, we do not consider  such  effect of the back shock.

 In Figure~\ref{fig:intl}, we can see that the maximum and minimum travel distances are located  at around the INFC and SUPC, respectively. For eccentric binary system (solid line in Figure~\ref{fig:intl}), we can see that the position of the maximum travel distance
slightly shifts from  INFC toward the apastron.  This is because the distance
between two stars, $D$,  changes along the orbital phase.  Figure~\ref{fig:intl} also indicates  that since  the variation of travel  distance along the orbit becomes  larger for the orbit with a  larger eccentricity,
the effect of the travel distance on the orbital modulation becomes larger  for the binary system with
more elongated orbit. In Figure~\ref{fig:intl1}, we can see that the amplitude of the variation of the travel distance along the orbit increases with increasing of the  momentum ratio, implying the effect could
be more important for the binary system with a  stronger  pulsar wind.
Actually, the effect of the travel distance  will  compensate with the effect
of the anisotropic field of the soft photon field of the ICS process, as we will discuss in later.

\subsection{Emission from the shocked pulsar wind.}
At the shock,  the kinetic energy of the pulsar wind before the shock is converted into the internal energy of the shocked pulsar wind. To evaluate the physical quantities of the pulsar wind just after the shock,
we apply the results of the  perpendicular  MHD shock \citep{1984ApJ...283..694K,1984ApJ...283..710K}.   The current approximation of the perpendicular shock  would be  valid for the emission happen only around the apex of the shock ($r\sim r_s$). A more realistic treatment would be more complicated, since the shock jump conditions depends on the inclination of the magnetic field relative to the shock surface and, the inclination varies with position of the shock region. In our model, however, the structure of the emission from the shock region are mainly originated around the apex. Hence, the current treatment could provide a good assumption to discuss the high-energy emission from the system. For the pulsar wind flow dominated by the particle kinetic energy,
that is,  low $\sigma$ limit ($\sigma << 1$), the radial four velocity   $u=\sqrt{\Gamma ^2-1}$ where $\Gamma$ is Lorentz factor of flow, proper number density,  and magnetic field strength  are expressed as
\begin{equation}
  u_2=\left(\frac{1+9\sigma}{8}\right)^{1/2}
\end{equation}
\begin{equation}
  n_2=\frac{n_1 u_1}{u_2}
\end{equation}
and
\begin{equation}
  B_2=3 B_1(1-4\sigma),
\end{equation}
respectively. Here the subscript  1 and 2 represent the physical quantities just before and after the shock, respectively. The number density and magnetic field strength just before the shock are calculated from
\begin{equation}
n_1 =\frac{\dot E_{sp}}{4\pi u_1 \Gamma_1 r_s ^2 m_e c^3 (1+\sigma)},
\end{equation}
and
\begin{equation}
B_1=\sqrt {\frac{E_{sp} \sigma}{r_s^2 c(1+\sigma)}},
\end{equation}
respectively,  where {\bf $\Gamma_1=\Gamma_W$} and $r_s$ is the distance to the shock from the pulsar.

For the down stream flow, the number  density and magnetic field is calculated from
the conservations that
\begin{equation}
  n(r)u(r)r^2=\mathrm{constant}
\end{equation}
\begin{equation}
  u(r)B(r)r/\Gamma_{sw}=\mathrm{constant}
\end{equation}
where $r$ is the distance from the shock and  $\Gamma_{sw}$ is the Lorentz factor of the shocked wind.
In this paper, the four velocity $u(r)$ is assumed to be a constant about 0.4 along the flow, that is, $\Gamma_{sw}\sim 1.08$, which is chosen to explain the orbital modulation of the observed X-ray emission.
 Here we assume that the Lorentz factor of post-shocked flow is constant. In \cite{2008MNRAS.387...63B} and their subsequent studies \citep{2019MNRAS.490.3601B},  they show that the post-shocked flow accelerates at the region far-from the apex ($r\gg r_s$), because of the adiabatic expansion. So our assumption of the constant speed may not be adequate except for around the apex. Based on our calculation, on the other hand, we would say that most of the emission from the shocked region is coming from the region around apex, where the magnetic field is stronger and the soft-photon energy density is larger. Therefore, the power of the emission far from the apex is much smaller than that around the apex. Hence the current assumption may be good enough.

We assume the initial distribution of the shocked particles follows a power law form,
\begin{equation}
  f_2(\gamma)\propto \gamma^{-p},~~\gamma_{min}\le \gamma\le \gamma_{max}
\end{equation}
and the normalization factor is determined from
\begin{equation}
  n_2=\int_{\gamma_{min}}^{\gamma_{max}}f_2(\gamma)d\gamma.
  \end{equation}
   The maximum Lorentz factor is determined by min$(\gamma_g,\gamma_{syn})$, where $\gamma_g$  is the Lorentz factor whose gyration radius is equal to the shock distance, and  $\gamma_{syn}$ is the Lorentz factor at which acceleration timescale is equal to the synchrotron cooling timescale,  $\tau_{ac}=\tau_{syn}$, where  $\tau_{ac}=\gamma m_e c/B e$ and $\tau_{syn} =9 m_e^3 c^5/e^4B^2\gamma$, respectively.  We assume $\gamma_{min}=4\times 10^5$ to explain the broadband spectrum.

\begin{figure}
  \includegraphics[scale=0.5]{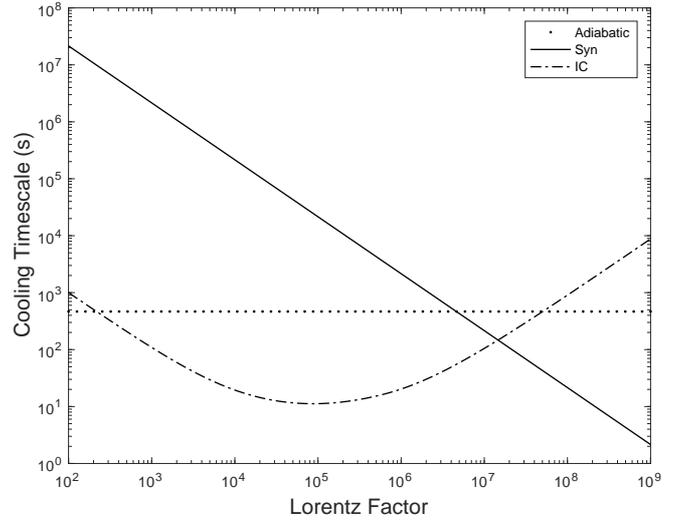}
  \caption{Cooling timescale at the periastron with the orbital parameters in Table~2.
    At the periastron, the separation of two stars is $D=0.25$AU and the magnetic field at the shock
    is about 0.6 Gauss.}
  \label{fig:cooling}
\end{figure}

 The particles  will loose  their energy owing to  the cooling processes;
adiabatic expansion, synchrotron emission, and IC scattering. We calculate the evolution of the Lorentz factor from
\begin{equation}
  \frac{d\gamma}{dt}=\frac{\gamma}{3n}\frac{dn}{dt}-\left(\frac{d\gamma}{dt}\right)_{syn}-\left(\frac{d\gamma}{dt}\right)_{IC},
\end{equation}
where the first term represents the  adiabatic expansion cooling. The synchrotron and ICS cooling processes are calculated from
\begin{equation}
  \left(\frac{d\gamma}{dt}\right)_{syn}=\frac{4e^4B^2{\gamma}^2}{9m_e^3c^5}
\end{equation}
and
\begin{equation}
  \left(\frac{d\gamma}{dt}\right)_{IC}=\int\int (E-E_s)\frac{\sigma_{IC}c}{m_e c^2E_s}\frac{dN_s}{dE_s}dE_sdE,
\end{equation}
respectively. Figure~\ref{fig:cooling} summaries the timescale of the cooling processes at the periastron (with the orbital parameters in Tables~1and~2) as a function of the Lorentz factor.

 From the number conservation in the phase space,
the distribution function at the distance $r$ can be calculated from
\begin{equation}
  f(r,\gamma)=\frac{n}{n_2}f_2\frac{d\gamma_2}{d\gamma},
\end{equation}
where the subscript 2 represents the initial value.

We calculate the ICS process of single particle with equation~(\ref{ics}). For the synchrotron radiation, the radiation power per unit energy is calculated
from
\begin{equation}
	P_{syn}=\frac{\sqrt{3}e^3 B\sin\theta_p}{h m_e c^2}F{\left(\frac{E}{E_c}\right)}
\end{equation}
$E_c=3he\gamma^2 B \sin\theta _p /4\pi m_e c$ is typical photon energy, $\theta_p$is the pitch angle and $F(x)=x\int _0 ^{\infty} K_{5/3}(y)dy$, and $K_{5/3}$ is the modified Bessel function. For the pitch angle,
we use the averaged value,  $\sin^2 \theta _p = 2/3$.

The Doppler boosting effect due to the finite speed of the post shocked flow enhances or suppresses  the observed emission and can cause  an  orbital modulation of the emission \citep{2010A&A...516A..18D,2014ApJ...790...18T}. The Doppler factor  is calculated  from
\begin{equation}
  D_{obs}=\frac{1}{\Gamma_{sw}(1-\beta_{sw} {\textbf e_{obs}} \cdot \textbf e_{flow})}
\end{equation}
where $\beta_{sw}$ is the speed of the shocked pulsar  wind in units of the speed of light, $\textbf e_{flow}$ is the unit vector along the direction of the flow and $\textbf e_{obs}$ is the unit vector in the direction of the observer.
To calculate the angle between the flow direction and observer direction,
we approximate that the flow is  in the orbital plane and moves away from the companion star, since the stellar wind pressure dominates the pulsar wind pressure ($\eta <<1$). The observed photon  energy and flux are   modified as $\epsilon= D_{obs} \epsilon ^{'}$ and $F_{\nu}(\epsilon)=D_{obs}^3F_{\nu} ^{'}(\epsilon^{'})$,  where the primed quantities refer to the value in the  co-moving frame.

\begin{figure}
	\includegraphics[scale=0.5]{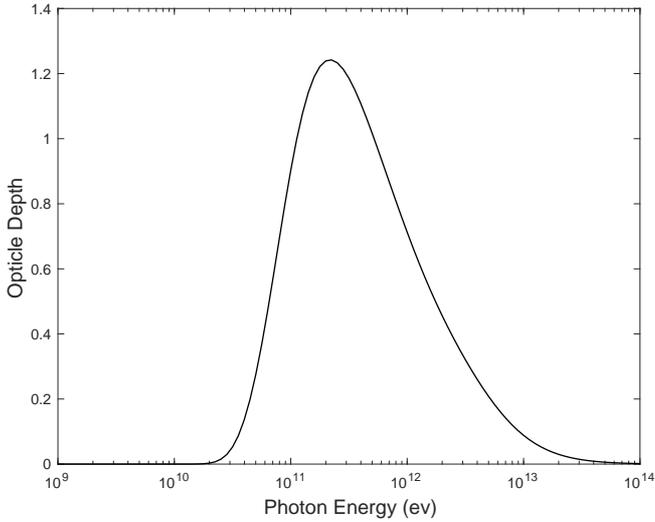}
	\caption{Optical depth along the line of sight of the photon-photon pair-creation process at the periastron of the binary orbit with the system inclination angle  $\alpha=30^{\circ}$ and eccentricity $e=0.4$.
	}
	\label{fig:opd}
\end{figure}

Finally, the pair-creation  process between the high-energy gamma-ray and the soft-photon from the companion star may affect to the observed TeV emission.
We calculate the  optical depth of the pair-creation process from \citep{1967PhRv..155.1404G}
\begin{equation}
	\tau_{\gamma\gamma} = \int _0 ^l dl \int _{4\pi} d\Omega(1-\mu)\int _{\frac{2}{\epsilon_{\gamma}(1-\mu)}}
	^{\infty} d\epsilon n_{ph}(\epsilon,\Omega)\sigma_{\gamma\gamma}
\end{equation}
where $l$ is the distance over which the $\gamma$-ray photon travels, $\mu = \cos \theta, d\Omega = d\mu d\phi$, and $n_{ph} (\epsilon ,\Omega)$ is the number density of the low-energy target photons. The cross section of the process is given as \citep{1976tper.book.....J},
\begin{equation}
 \sigma_{\gamma\gamma}=\frac{3}{16}\sigma_T(1-\beta^2)\left[(1-\beta^4)\mathrm{ln}\left(\frac{1-\beta}{1+\beta}\right)-2\beta(2-\beta^2)\right]
\end{equation}
where
\begin{equation}
	\beta=\sqrt{1-\frac{1}{\epsilon\epsilon_{\gamma}(1-\mu)}}
\end{equation}
and $\sigma_T$ is the Thomson cross section. Figure~\ref{fig:opd} shows the optical depth  of the photon traveling toward the observer at periastron.  For the photons with an energy less than $10^{10}$eV, the pair-creation process is negligible,  as shown in the figure.

\section{Result}
\label{obser}
In this section, first  we will  discuss on the GeV emission within the framework of the ICS model of the unshocked pulsar wind and discuss the dependency of the shape of the light curve on the system parameters. Then, we will compare the model light curve with the Fermi-$LAT$ observation of the gamma-ray binary LMC~P3.  We compare  the predicted X-ray and TeV emissions from the shocked pulsar wind  with the observations for LMC~P3. We discuss the system geometry of 4FGL~J1405.4-6119 by fitting the observed GeV light curve.

\begin{table*}
  \centering
  \caption{Parameters of the pulsar and O star applied in this study.}
  \begin{tabular}{ccccc}
    \hline
    Spin-down power & Magnetization parameter & O star radius & O star temperature  & Wind momentum ratio\\
 $L_{sp}$ & $\sigma$  & $R_{c}$ & $T_c$ & $\eta$ \\
\hline
 $6\cdot 10^{36}$erg/s   &0.003 & $14.5 R_{\odot}$ & $2.5\cdot10^4$K & 0.01\\
    \hline
    \end{tabular}
\end{table*}

\begin{table*}
  \centering
  \caption{Parameters of the systems applied in Figure~\ref{fig:lcumev2}.}
  \begin{tabular}{cccccc}
    \hline
 Orbital period &   Eccentricity & Inclination angle  & Periastron & INFC &SUPC \\
   \hline
10.3{\rm days} & 0.4 & $30^{\circ}$ &  0.03& 0.24&0.98 \\
    \hline
    \end{tabular}
\end{table*}

\begin{figure}
  \includegraphics[scale=0.5]{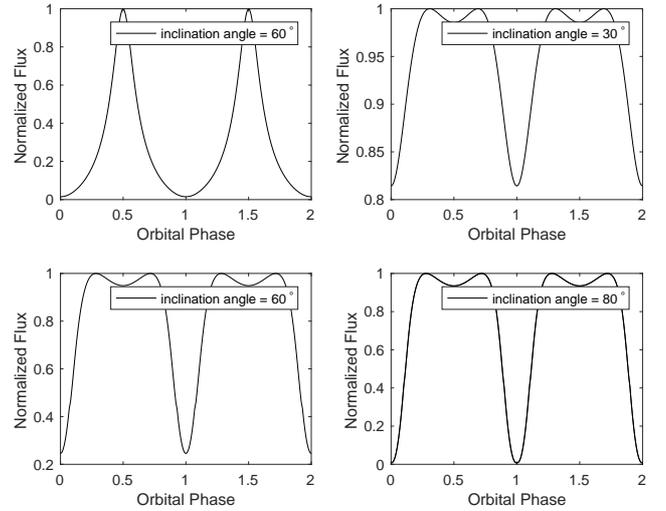}
  \caption{Dependency of the orbital modulation of ICS emission from unshocked pulsar wind with the system inclination angle. The result is for the circular orbit and $\eta=0.03$. In the top left panel, the orbital modulation by assuming
    the constant travel distance of the pulsar wind is presented for the comparison.  The INFC and SUPC happen $\Phi _{INFC}=0$ and $\Phi_{SUPC}=0.5$, respectively }
  \label{fig:errb3}
\end{figure}

\begin{figure}
  \includegraphics[scale=0.5]{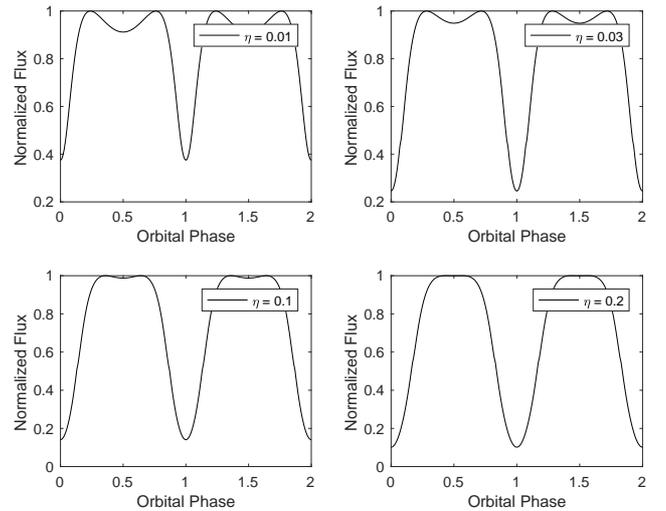}
  \caption{Dependency of the orbital modulation of ICS emission from unshocked pulsar wind with the momentum ratio of two winds. The result is for the circular orbit and the system inclination angle $\alpha=60^{\circ}$. The INFC and SUPC happen {\bf at} $\Phi _{INFC}=0$
    and $\Phi_{SUPC}=0.5$, respectively }
  \label{fig:errb7}
\end{figure}

\subsection{ICS emission from  unshocked pulsar wind}
\label{lgev}

We assume that the GeV emission from the gamma-ray binary is produced via ICS 
process of the unshocked pulsar wind and assume that the particles in the unshocked pulsar wind have a mono-energetic distribution (in this paper, we assume the Lorentz factor to be $3\times 10^4$, c.f. equation~(\ref{gammaw})). In section~3.1, we discussed a detailed of the calculation method. The variability of the emissivity of ICS process along the  orbital phase is mainly determined by the variability of three effects, namely, (i) the size of the emission region, (ii) collision angle between the pulsar wind and soft-photon from the companion star,  and (iii) soft photon and the density of background photon. These parameters then tightly relate to the system parameters, momentum ratio of the two winds,   inclination angle of the system,  eccentricity, and the true anomaly of the INFC/SUPC measure from the periastron. In this section, therefore, we investigate how orbital variation depends on the system parameters.

Firstly, Figure~\ref{fig:errb3} summarizes how the inclination angle of the system affects to the light curve by using the circular orbit. For the circular orbit, since  the distance to the apex of the shock from the pulsar is constant with the orbital phase, the soft photon density at the unshocked pulsar wind region is also almost constant. Hence, the orbital variation is mainly caused by the effects of the collision  angle and the travel length, which depends on the inclination angle of the system.  In Figure~\ref{fig:errb3}, we set the momentum ratio to be $\eta=0.03$ and, the INFC and SUPC happen at $\Phi_{INF}=0$ and $\Phi_{SUPC}=0.5$, respectively.  The top-right, bottom-left and bottom-right panels show the result for $\alpha=30^{\circ}$, $60^{\circ}$ and $80^{\circ}$, respectively. In the  top left column, we present the orbital modulation with $\alpha=60^{\circ}$ by assuming the constant travel distance of unshocked pulsar wind for the comparison (i.e., the orbital modulation is only affected by the variation  of the collision angle).  For a constant travel length with the orbital phase, the observed flux acquire  the  maximum value at the SUPC and the minimum value at the INFC, since the collision angle is the maximum (head-on)  and minimum (tail-on) at the SUPC and INFC, respectively. 

As Figure~\ref{fig:errb3} shows, we find that  the  overall feature of the light curve does not change, even the effect of the travel distance is taken into account.  However we find that the light curve has a small dip at SUPC. This is because the shortest travel distance occurs at SUPC, as discussed in section~3. By the comparing the light curves between top-left and bottom-left panels for $\alpha=60^{\circ}$, we can find that the effect of the travel length (i) reduces the amplitude of the orbital variation and (ii) makes the width of the peak wider. 
By  comparing among  top right, bottom-left  and bottom-right panels, we see that as the inclination angle increases, the amplitude of  variation becomes larger. This suggests that  the effect of travel length become more significant for a smaller inclination angle. 

In Figure~\ref{fig:errb7}, we  summarize  the dependency of the momentum ratio $\eta$ ($=L_{sd}/(\dot{M}v_wc)$ for the circular orbit $e=0$. The momentum ratio affects to the variation of the observed flux, since it affects to  (i) the  travel length of the unshocked pulsar wind (c.f. Figure~\ref{fig:intl1}) and (ii) the density soft-photon field at the emission region. For example,  Figure~\ref{fig:intl1} shows that for larger momentum ratio,  the ratio of the travel distance at the SUPC to  one  at the INFC is smaller, and therefore the dip at SUPC tends to be dipper.  The  second effect about the soft photon density  is caused  because for the larger momentum ratio, the apex of the shock is located at the position closer to the companion star, and the soft photon number density of the emission region become larger.   This effect tends to increase the emissivity at around the SUPC, since the  pulsar wind approaching toward the companion star produces the observed emission at around the SUPC.   From Figure~\ref{fig:errb7}, we find that  the dip at the SUPC becomes shallow   with increase of the momentum ratio (stronger pulsar wind). This is because the second  effect of the photon density overcomes the first effect of the travel length. 
The current calculation suggests therefore 
that the effect of the travel distance is more important for the binary system in which the stellar wind 
is much stronger than that of the pulsar wind.

\begin{figure}
        \includegraphics[scale=0.5]{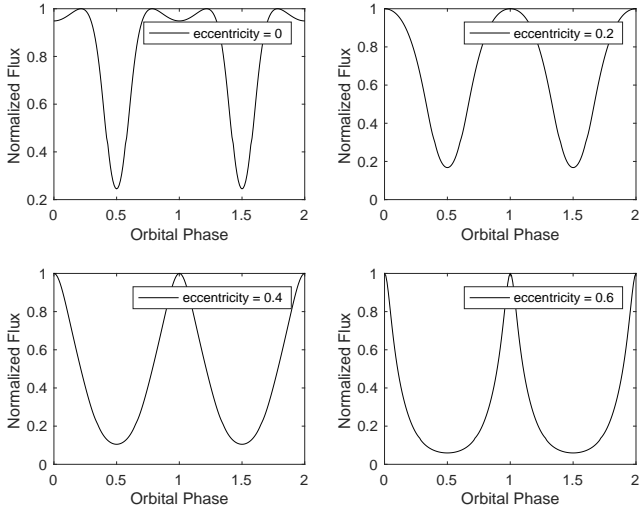}
        \caption{Dependency of the orbital modulation of ICS emission from unshocked pulsar wind with the eccentricity. The result is for inclination angle is 60 degrees,.The INFC and SUPC happen {\bf at} $\Phi _{INFC}=0.5$  and $\Phi_{SUPC}=0$, respectively . }
        \label{fig:errb1}
\end{figure}

\begin{figure}
        \includegraphics[scale=0.5]{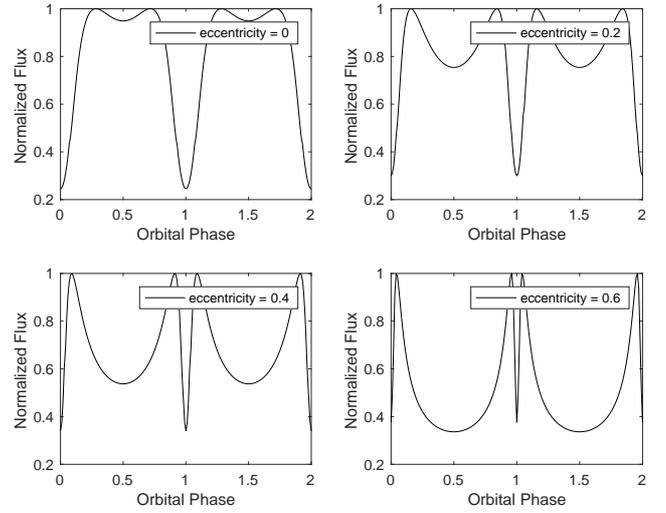}
        \caption{Dependency of the orbital modulation of ICS emission from unshocked pulsar wind with the eccentricity. The result is for the system inclination angle $\alpha=60^{\circ}$.  The INFC and SUPC happen at $\Phi _{INFC}=0$ and $\Phi_{SUPC}=0.5$ .}
        \label{fig:errb2}
\end{figure}

\begin{figure}
        \includegraphics[scale=0.5]{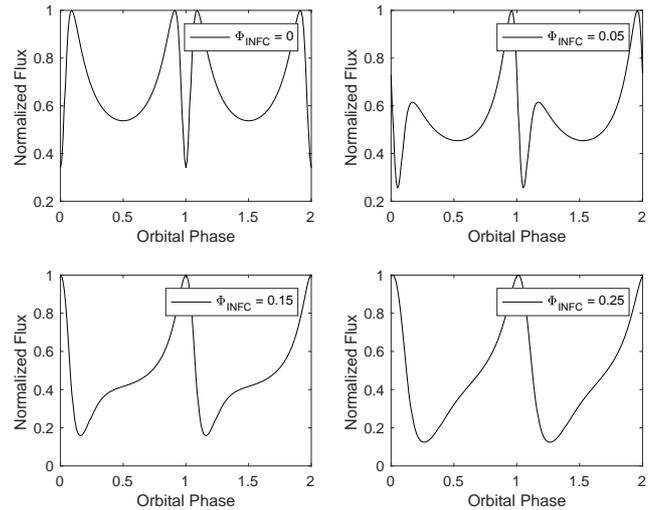}
        \caption{Dependency of the orbital modulation of ICS emission from unshocked pulsar wind with the position of INFC.The result is for eccentricity is 0.4.  The position of INFC and SUPC are
          ($\Phi_{INFC},~\Phi_{SUPC}$)=(0,~0.5) for upper-left panel, (0.05,~0.73) for upper-right panel,
          (0.15~0.89) for lower-left panel and (0.25,~0.94), respectively. The periastron is located at
          $\Phi_{peri}=0.0$ in each panel. }
        \label{fig:errb6}
\end{figure}

In Figures~\ref{fig:errb1}-\ref{fig:errb6}, we discuss the  dependency of the light curve on (i)  the eccentricity and (ii) the position of the INFC/SUPC relative to the periastron.  The eccentricity introduces an  variation of the distance to the apex from the pulsar along the orbital phase, and the periastron (or apastron) is  defined at the position where  the distance between the pulsar and companion star is the shortest (or the  longest).  

In Figure~\ref{fig:errb1}, to investigate the dependency of the eccentricity, we assume that the position of periastron and apastron are coincide with  the SUPC and  INFC, respectively, and we choose $\Phi_{peri}=\Phi_{SUPC}=0$ and  $\Phi_{apa}=\Phi_{INFC}=0.5$.  From  Figures~\ref{fig:errb1}, we see  that as the eccentricity increases, 
the double peak structure disappears and the peak at periastron/SUPC becomes sharper. This can be understood because for a larger eccentricity, the distance between the pulsar and companion star at the periastron is shorter and hence the density of soft-photon at the emission region is larger. As Figure 3 shows, with increase of the eccentricity, (i) the travel distance at the SUPC decreases and (ii) the variation of the amplitude along the orbit increases. Although this effect of the travel lengths  tends to decrease the observed flux at the SUPC, it is overcame by the effect of the increase in the soft-photon density at the emission region. As a result, the model light curve has a more prominent peak at SUPC for
 larger eccentricity. We conclude therefore that the effect of the travel distance is more important 
 for the pulsar binary system with a lower eccentricity.

In Figure~\ref{fig:errb2}, we consider the opposite case of Figure~\ref{fig:errb1}, namely, we assume the  position of the periastron and apastron  at  $\Phi_{peri}=\Phi_{INFC}=0$ and  $\Phi_{apa}=\Phi_{SUPC}=0.5$, respectively. In this case, we can see local minima located  at SUPC and INFC for all eccentricity.
The local minimum at apastron/SUPC is created as a result of the minimum of the soft-photon density at the emission region, and the minimum at periastron/INFC is produced as a result of the minimum of the collision angle. 

 Since 4FGL J1405.4-6119 likely shows the double peak structure in the GeV light curve, its position of the INFC/SUPC relative to periastron/apastron  could be similar to  the case of Figure~\ref{fig:errb2}. In Figure~\ref{fig:errb6}, therefore, we summarizes how the shape of the  light curve depends on the periastron/apastron relative to the SUPC/INFC with  the eccentricity $e=0.4$ and the system inclination angle $\alpha=60^{\circ}$.  In the top-left panel, we represent the double peak light curve with the geometry $\Phi_{INFC}=\Phi_{peri}=0$.
 Then we shift the position of the INFC to $\Phi_{INFC}=0.05$ (top-right), 0.15 (bottom-left) and 0.25 (bottom-right), respectively; the position of the periastron and apastron are fixed  at $\Phi_{peri}=0$ and $\Phi_{apa}=0.5$, respectively.  As the position of the INFC, at which the calculated flux is minimum, is shifted away from the periastron  (in other words, the SUPC is shifted toward the periastron), the flux minimum also shifts, and  the dip appeared at SUPC becomes shallower. The calculated light  curve  is eventually described by the single peak.

We created  the  light curve with the various system parameters  to examine allowed range of the parameters that produce  the double peak structure.   For the eccentricity $e=0.4$ and the inclination angle $\alpha=60^{\circ}$ of Figure~\ref{fig:errb6}, for example,  we find that the allowed range of the position of INFC that create the double peak structure in the light curve is $\delta|\Phi_{INFC}|< 0.05$ measured from the periastron. We note that although the allowed range if $\Phi_{INFC}$ is within 10\% of the orbital phase measured in time, it corresponds to $\le 50^{\circ}$ in the true anomaly for the eccentricity $e=0.4$. Hence, such a viewing geometry will be not uncommon. We can find that the allowed range of the $\Phi_{INFC}$ in true anomaly is not sensitive to the eccentricity, but a larger eccentricity creates double peaks with a smaller phase separation, as indicated in Figure~\ref{fig:errb2}.  As decrease of the system inclination angle from $\alpha=60^{\circ}$, the double peak structure transits to the single peak. With the eccentricity $e=0.4$ and $\Phi_{INC}=0.05$ (tot-right panel in Figure~\ref{fig:errb2}, for example,  the lower limit of the inclination angle to create the double peak structure is $\sim 40^{\circ}$. 

\subsection{Application to LMC~P3}

To explain the observed flux level, \cite{2016ApJ...829..105C} suggests that the spin-down power of the  pulsar is about
$L_{sd}=4.3\times 10^{36}{\rm erg~s^{-1}}$. In this study, we assume $L_{sd}=6\times 10^{36}{\rm erg~s^{-1}}$. To estimate the magnetic field at the shock, we apply  $\sigma=0.003$, which is measured for the Crab pulsar,  of the magnetization parameter at the termination shock. With $\sigma=0.003$, the magnetic field strength at the shock is of\
the order of $1$G.   We also assume  $R_c=14.5 R_{\odot}$ for the radius of companion star to estimate the
angular size of the companion star, $\theta_{c}$, in the equation of (\ref{ics}), and assume
  $T_{c}\sim 2.5\times 10^4$K for the temperature of the companion star.
  Tables~1 and 2 summarize the parameters of the system applied in this study. The pulsar goes around the O star in the orbit shown in figure\ref{fig:obt}.

\subsubsection{GeV light curve}
The current model assumes that the GeV emission modulating with the orbital phase is produced by the ICS of the unshocked pulsar wind off the stellar photon. The GeV emission from LMC~P3 gamma-ray binary shows  a  broad asymmetric peak structure and probably has a  double peak structure in the light curve.  As we  discussed above, the shape of the light curve depends on the system parameters, $\eta$, $\alpha$ and $e$.  To fit the observed radial velocity curve, \cite{2016ApJ...829..105C}    obtain the SUPC at the $\Phi_{SUPC}\sim 0.8-0.9$, where the phase zero is defined  at MJD~57,410.25. \cite{2019MNRAS.484.4347V} refine  the orbit parameter with a more detailed optical observation. With $\Phi_{orb}=0$ at MJD~57,410.25, the periastron,  INFC and SUPC occur at $\Phi_{peri}=0.13$, $\Phi_{INFC}=0.24$ and  $\Phi_{SUPC}=0.98$, respectively. The eccentricity is measured as $e\sim 0.4$,  and the measured mass function implies that  the inclination angle, $\alpha$, is between  $\alpha \sim 40^{\circ}$ and $60^{\circ}$ with $1.4M_{\odot}$  for the mass of the compact object and $25-40M_{\odot}$ for the mass of companion star.

\begin{figure}
        \includegraphics[scale=0.5]{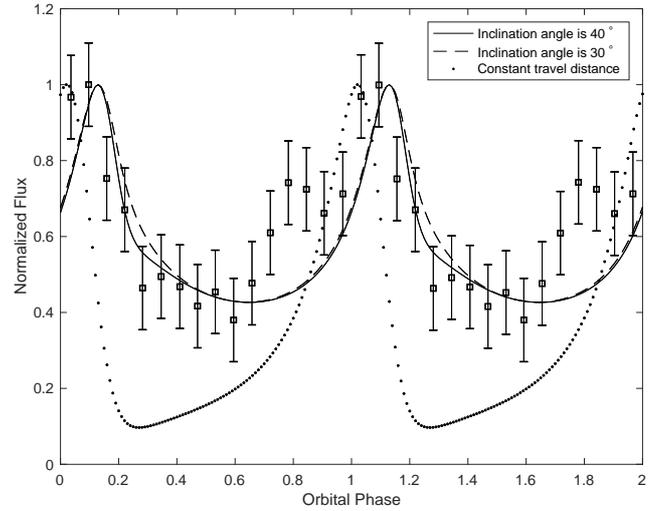}
        \caption{GeV light curve of LMC~P3. The lines are model light curve with the system
 parameters suggested in the \citealt{2019MNRAS.484.4347V}.  The periastron, INFC and SPUC are $\Phi_{peri}=0.13$,$\Phi_{INFC}=0.24$ and $\Phi_{SUPC}=0.98$,  respectively. And the inclination angle is $40^{\circ}$ for solid line and dash line for $30^{\circ}$. The data are taken from \citealt{2016ApJ...829..105C}}
        \label{fig:lcumev1}
\end{figure}

Figure~\ref{fig:lcumev1} compares the light curves observed by $Fermi$-LAT ($>100$MeV) with the  calculated
light curve with using the systems parameters obtained by  \cite{2019MNRAS.484.4347V}; we assume the momentum ratio  $\eta=0.03$ and the inclination angle  $\alpha \sim 40^{\circ}$ which is lower limit for the neutron star  mass 1.4$M_{\odot}$ given  by \cite{2019MNRAS.484.4347V}.
The dotted line and solid line in the figure are the calculated light curve without and with the effect of the travel distance of the unshocked pulsar wind, respectively. We  find  in the figure that the modulation in the calculated light curve ignoring the effect of the travel distance becomes significantly larger that that of the the observations. We also see that the width of the peak  located at the SUPC is narrower than the observation. By taking account for the effect of the travel distance (solid line in the figure), we can see in the figure that the amplitude of the calculated light curve (solid line) is reduced and it is more consistent with the observation. With the suggested orbital parameters, therefore, the effect of the travel distance of the unshocked pulsar wind is important to explain the observed orbital modulation, proving that the GeV emission is originated from the unshocked pulsar wind.

\begin{figure}
        \includegraphics[scale=0.5]{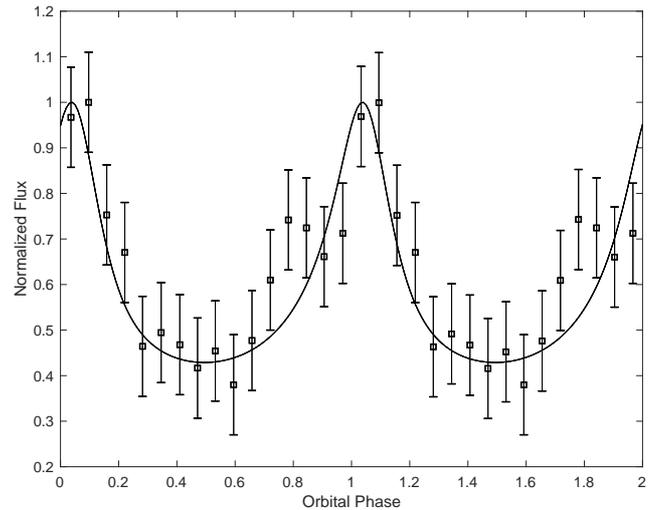}
        \caption{GeV light curve of LMC~P3. The solid line is model light curve with the system
 parameters presented in the Table~1. The data are taken from \citealt{2016ApJ...829..105C}}
        \label{fig:lcumev2}
\end{figure}

As the solid line in Figure~\ref{fig:lcumev1} shows, the model light curve with the inclination angle $\alpha=40^{\circ}$ would have a peak narrower than the observed one and the model flux at $\Phi_{orb}\sim 0.8$ would be significantly deviated from the observation, although the uncertainties of the error are large. We therefore calculate the model light curve with different inclination angle, and we find that the model light curve with a different inclination angle does  not improve this feature.
In Figure~\ref{fig:lcumev1}, for example, we present the model light curve  with the inclination angle $\alpha=30^{\circ}$,
for which the mass of the compact object is slightly larger than 1.4$M_{\odot}$ with $25-40M_{\odot}$ of a  companion star.
Due to the observed uncertainty of orbital parameters, we shift the position of the SUPC, periastron  and INFC to  $\Phi_{orb}=0.98, 0.03$ and 0.24, respectively, which are  still in the range suggested by \cite{2019MNRAS.484.4347V}.
Figure~\ref{fig:lcumev2} compares the model light curve with the observations.  Compared with the model light curves (soled and dashed lines) in Figure\ref{fig:lcumev1}, the position of peak is shifted because of the applied positions of SUPC and periastron. The feature of slow rising  and rapid decreasing of the peak shape and the peak width would be more consistent with the observations within the range of the errors. We note that current model does not expect the double peak structure of the light curve, as seen in Figure~\ref{fig:errb6}, with the system parameters of the LMC~P3, because we observe the system from the direction far from the periastron.

\subsubsection{X-ray/TeV light curves}
We assume that the  synchrotron emission and ICS process of the shocked pulsar wind
particles  produce the X-ray and TeV gamma-ray, respectively, of LMC~P3. Using the system parameters
obtained from the fitting of the GeV light curve in Figure~\ref{fig:lcumev2} (Table~2),
we calculate the shock emission.

\begin{figure}
        \includegraphics[scale=0.5]{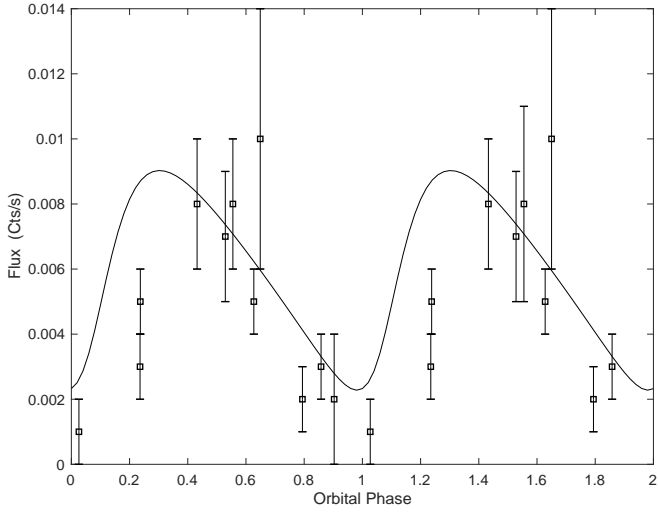}
        \caption{X-ray light curve of LMC~P3. The solid line is model light curve with the system
 parameters presented in the Table~1. The data are taken from \citealt{2016ApJ...829..105C}}
        \label{fig:lcukev}
\end{figure}

Figure~\ref{fig:lcukev} compares the observed X-ray light curves with calculated X-ray light curve.
In the current model, the orbital modulation of the X-ray is caused by (i)
the Doppler boosting process \citep{2010A&A...516A..18D} and (ii) variation of
the shock distance from the pulsar/companion star, which causes the variation of the cooling
timescale along the orbit \citep{2007Ap&SS.309..261K,2009ApJ...702..100T}.
In the current emission  model, we find that the second effect on the orbital modulation
cannot reproduce the observed amplitude of the orbital variation in the X-ray bands,
and therefore we expect that  the Doppler boosting effect mainly causes  the observed orbital variation.
With the Doppler boosting effect,
the maximum and minimum intensity of the X-ray emission tend to appear
 at INFC and SUPC, respectively. To explain the amplitude of the observed modulation, we
 assume the Lorentz factor of the shocked pulsar wind in the value of
 $\Gamma_{sw}\sim 1.08~(\beta_{sw}\sim 0.35$).

\begin{figure}
	\includegraphics[scale=0.5]{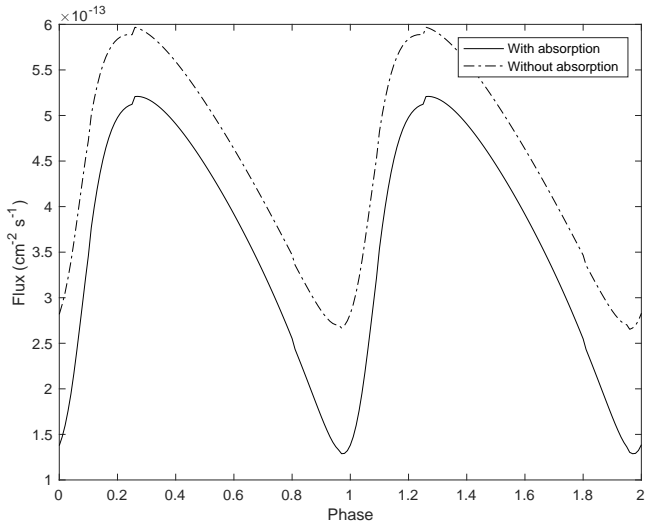}
	\caption{The predicted light curve in 1-100TeV energy bands with the system parameters of
          Tables~1 and~2. The solid line and dashed line represent the model light curves with absorption and without absorption respectively.}
	\label{fig:lct}
\end{figure}

\begin{figure}
  \includegraphics[scale=0.5]{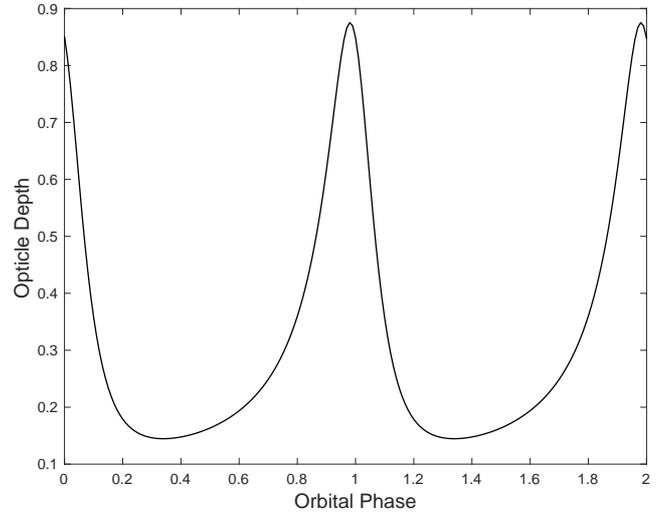}
  \caption{The orbital variation of optical depth along the line of sight for 1~TeV photon with the system parameters of Tables~1 and~2.}
  \label{opd12}
    \end{figure}

Figure \ref{fig:lct} presents the predicted  TeV light curve. In the TeV energy bands, the orbital
modulation is caused by (i) the Doppler boosting, (ii) the variation of the soft-photon field at the emission
region and (iii) the absorption owing to the pair-creation process. With the assumed system parameters in Table~1 and~2, we find that optical depth of 1TeV photons that are  traveling toward the observer is less than unity for whole orbit (Figure~\ref{opd12}).
With the inclination angle $\alpha=30^{\circ}$, the effect of the collision angle on the orbital variation is less important, and
therefore the Doppler boosting effect dominates in the orbital modulation of the model light curve.
Our model predicts that the  light curve has a peak (or minimum) intensity around  the INFC (or SUPC). \cite{2018A&A...610L..17H} found a significant TeV emission around the INFC, which is consistent with the current model. Since the TeV emission at most of orbital phase  has not been confirmed by the current observation, we do not pursue a detail comparison  between the model and observed light curves.


\subsubsection{Broadband spectrum}

\begin{figure}
	\includegraphics[scale=0.5]{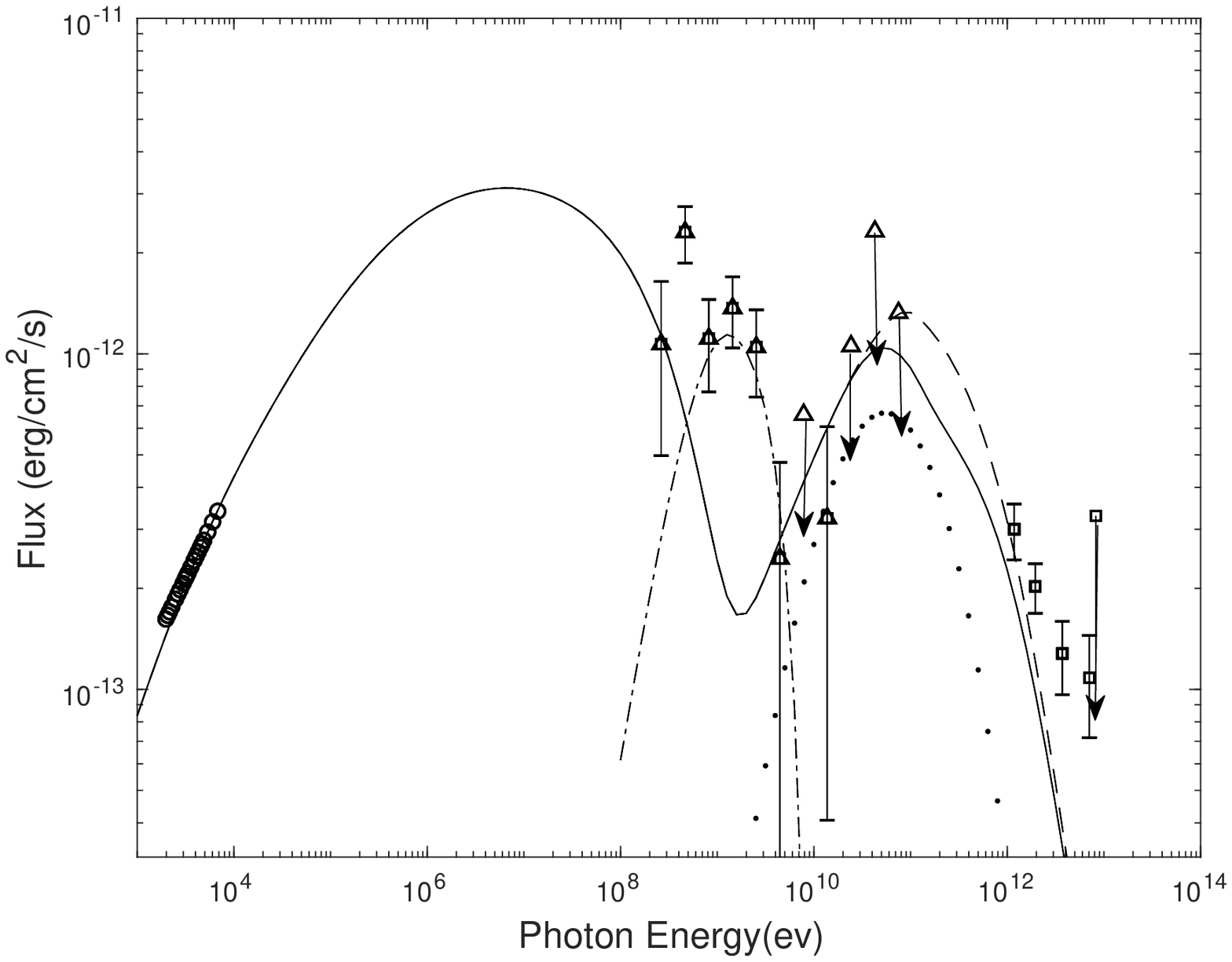}
	\caption{ The solid line is spectra of shock emission with photon-photon absorption. The dash line is the spectrum without the  absorption. Dash-dot line is PW emission and the dot line is secondary emission. X-ray  data and TeV data are  taken from \citealt{2016ApJ...829..105C} and
          from \citealt{2017heas.confE..33K}, respectively.}
	\label{fig:spc}
\end{figure}

Figure~\ref{fig:spc} compares between the calculated and observed spectra averaged over the whole orbit.
In the current model, the ICS of the unshocked pulsar wind (dotted-dashed line) explains
the observed spectrum  in $10^{8-9}$eV energy bands, while the synchrotron radiation and ICS process
of the shocked pulsar wind (solid line) explain the observed X-ray and TeV emission, respectively.

It has been discussed the emission from  the secondary pairs created at the stellar wind region
 by the photon-photon annihilation  process
 \citep{1997A&A...322..523B,2008MNRAS.385.2279S,2007ApJ...671L.145S,2010ASPC..422...41C}.   As shown  in Figure \ref{fig:opd}, the photon with an  energy larger than $\sim 10^{10}$eV may be
 converted into the pairs, and therefore  the Lorentz factor of the secondary
 are typically in the range of $\sim 10^5-10^7$.  To estimate the synchrotron emission
of the secondary pairs created  at the stellar wind region,
we assume that  the magnetic field where the  secondary emission occurs is dominated by O star's magnetic
field and  we apply a simple power law form $B(R)=B_c(R/R_c)^{-m}$ with $B_c$ being the stellar magnetic field and $R$ distance to the emission region
from the star. The stellar magnetic field of the high mass main-sequence star can be $B_c\sim 10^{2-3}$G \citep{2012SSRv..166..145W} . With a typical value
$m=2-3$, we can see that the synchrotron emission of the second pairs is stronger than ICS process only at near the companion star. In the current
model, therefore, the synchrotron emission from the secondary is negligible in the observed
emissions. To discuss the ICS process, we assume that
the secondary pairs created in the stellar wind region  are quickly isotropized and
we calculate the ICS  process with   a constant soft-photon field during the crossing timescale $D/c$.  In Figure~\ref{fig:spc}, the dotted line  represents the contribution of the emission from the secondary pairs.

\subsection{Application to 4FGL J1405.4-6119}
\begin{figure}
  \includegraphics[scale=0.5]{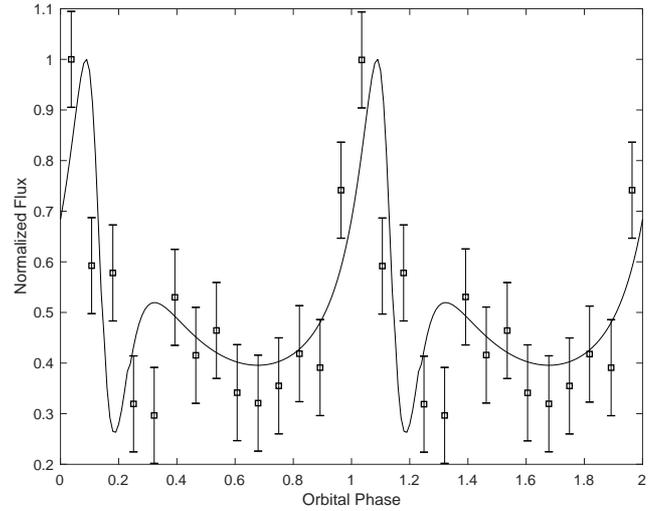}
  \caption{GeV light curve of 4FGL J1405.4-6119.  The solid line is model light curve with
    the orbit parameters $\eta=0.03$, $\alpha= 60^{\circ}$ and $e=0.4$.
    We assume $\Phi_{peri}=0.1$,  $\Phi_{INFC}=0.18$ and $\Phi_{SUPC}=0.91$, and apply
    the parameters of the companion star in Table~1. The data points of the $Fermi$-LAT are taken from \citealt{corbet19}. }
  \label{fig:J1405}
\end{figure}

4FGL J1405.4-6119 is the new gamma-ray binary discovered by $Fermi$ \citep{corbet19}.
The GeV light curve measured by $Fermi$ would show a double peak structure with the small peak and sharp strong peak (Figure~\ref{fig:J1405}). The strong peak is probably shifted from the peak of the X-ray \citep{corbet19}, which is similar to  LMC~P3 and LS~5039. We therefore expect that GeV emission from the source is originated from the ICS of the unshocked pulsar wind.

The system parameters have not been determined yet.
In this paper, therefore, we assume  the eccentricity $e=0.4$, the momentum  ratio $=0.03$ and the temperature of the companion star  $T_c\sim 2.5\times 10^4$K, which were applied for the case of LMC~P3. The expected phase of the periastron, SUPC and INFC are determined by the fitting of the GeV light curve, as shown in
Figure~\ref{fig:J1405}. As we discussed in section~\ref{lgev}, the current model predicts the double peak structure if the SUPC  happens at around apastron, namely, we observe the system from the direction of the periastron. In Figure~\ref{fig:J1405}, we fit the $Fermi$ data with $\Phi_{peri}=0.1$, $\Phi_{INFC}=0.18$ and $\Phi_{SUPC}=0.91$; the true anomalies of INFC and SUPC
measured from the periastron are $65^{\circ}$ and $245^{\circ}$, respectively. Our model predicts that the sharp peak is located at around the periastron, and there is a local minimum at the phase between the apastron and SUPC. The requirement for reproducing the double peak structure that we observe the system from the direction of the periastron is not sensitive to the eccentricity and the system inclination angle.
The future observations for the system parameter will be compared with  the model prediction.

\section{Summary}
\label{summary}
We have studied the GeV emission from the gamma-ray binary systems  composed of young pulsar and O-type main-sequence star. In our model, the GeV emission is originated from the ICS process of the unshocked
pulsar wind off the soft-photons from the companion star. The unshocked pulsar wind is stopped by
the intra-binary  shock and the travel distance of the unshocked wind that moves toward the observer
depends on the orbital phase. In this paper, we studied how the effect of the travel distance affects to the orbital modulation of the observed GeV emission.

Comparing with the model light curve in which the constant travel distance along the orbital phase is assumed, the effect of the travel distance tends to create a small dip around the SUPC, where the travel distance is the minimum. In addition, the effect  makes the peak width wider. In a real situation, the effect of the travel distance compensates with the effect of anisotropic soft-photon field, and this effect is more prominent for the binary system with a lower eccentricity and a stellar wind being much stronger than the pulsar wind.
For the system with the higher eccentricity, the shape of the light curve and
the peak position are  mainly affected by the effect of the anisotropic soft-photon field.

We apply the our model to two  gamma-ray binaries, LMC~P3 and 4FGL J1405.4-6119 that  were recently discovered by $Fermi$-LAT. Applying the system parameters of LMC~P3 suggested by \cite{2019MNRAS.484.4347V}, our model light curve with the effect of the travel distance would be consistent with the observed GeV light curve. In particular, we found that the observed  amplitude of the modulation and width of the peak can be described as a  result of the effect of the travel distance. We also calculated the X-ray/TeV emission from the shocked pulsar wind and found that  the Doppler boosting effect is more important to explain the observed X-ray modulation with the suggested system parameters. Within the current model, we suggest that the observed double peak structure in the GeV light curve of 4FGL J1405.4-6119 is owing to the anisotropic soft-photon field and we predict that the system is viewed from the direction close to  the periastron.  Future observations for the system parameters will be compared with the our  prediction to constrain the GeV emission process.

 We wish to express our thanks to the referee for detailed comments to  improved the paper. We also thank to 
Prof. K.S. Cheng and Dr. A.M. Chen for useful discussion for the gamma-ray binaries. 
 This work has made use of data supplied by the Fermi Science Support Center and the software supplied by the Fermipy Developers. H.X.X. and J.T.  are  supported by NSFC grants of the Chinese Government under 11573010, 11661161010, U1631103 and U1838102. T.Q.W. is supported by the NSFC under 11903017 and 11547029.

\bibliography{mndraft}

\begin{thebibliography}{}
\makeatletter
\relax
\def\mn@urlcharsother{\let\do\@makeother \do\$\do\&\do\#\do\^\do\_\do\%\do\~}
\def\mn@doi{\begingroup\mn@urlcharsother \@ifnextchar [ {\mn@doi@}
  {\mn@doi@[]}}
\def\mn@doi@[#1]#2{\def\@tempa{#1}\ifx\@tempa\@empty \href
  {http://dx.doi.org/#2} {doi:#2}\else \href {http://dx.doi.org/#2} {#1}\fi
  \endgroup}
\def\mn@eprint#1#2{\mn@eprint@#1:#2::\@nil}
\def\mn@eprint@arXiv#1{\href {http://arxiv.org/abs/#1} {{\tt arXiv:#1}}}
\def\mn@eprint@dblp#1{\href {http://dblp.uni-trier.de/rec/bibtex/#1.xml}
  {dblp:#1}}
\def\mn@eprint@#1:#2:#3:#4\@nil{\def\@tempa {#1}\def\@tempb {#2}\def\@tempc
  {#3}\ifx \@tempc \@empty \let \@tempc \@tempb \let \@tempb \@tempa \fi \ifx
  \@tempb \@empty \def\@tempb {arXiv}\fi \@ifundefined
  {mn@eprint@\@tempb}{\@tempb:\@tempc}{\expandafter \expandafter \csname
  mn@eprint@\@tempb\endcsname \expandafter{\@tempc}}}

\bibitem[\protect\citeauthoryear{{Abdo} et~al.,}{{Abdo}
  et~al.}{2009}]{2009ApJ...706L..56A}
{Abdo} A.~A.,  et~al., 2009, \mn@doi [\apjl] {10.1088/0004-637X/706/1/L56},
  \href {https://ui.adsabs.harvard.edu/abs/2009ApJ...706L..56A} {706, L56}

\bibitem[\protect\citeauthoryear{{Abdo} et~al.,}{{Abdo} et~al.}{2011}]{abdo11}
{Abdo} A.~A.,  et~al., 2011, \mn@doi [\apjl] {10.1088/2041-8205/736/1/L11},
  \href {https://ui.adsabs.harvard.edu/abs/2011ApJ...736L..11A} {736, L11}

\bibitem[\protect\citeauthoryear{{Abeysekara} et~al.,}{{Abeysekara}
  et~al.}{2018}]{2018ApJ...867L..19A}
{Abeysekara} A.~U.,  et~al., 2018, \mn@doi [\apjl] {10.3847/2041-8213/aae70e},
  \href {https://ui.adsabs.harvard.edu/abs/2018ApJ...867L..19A} {867, L19}

\bibitem[\protect\citeauthoryear{{Ackermann} et~al.,}{{Ackermann}
  et~al.}{2016}]{2016A&A...586A..71A}
{Ackermann} M.,  et~al., 2016, \mn@doi [\aap] {10.1051/0004-6361/201526920},
  \href {https://ui.adsabs.harvard.edu/abs/2016A%26A...586A..71A} {586, A71}

\bibitem[\protect\citeauthoryear{{Aharonian} et~al.,}{{Aharonian}
  et~al.}{2005}]{2005A&A...442....1A}
{Aharonian} F.,  et~al., 2005, \mn@doi [\aap] {10.1051/0004-6361:20052983},
  \href {https://ui.adsabs.harvard.edu/abs/2005A&A...442....1A} {442, 1}

\bibitem[\protect\citeauthoryear{{Aharonian} et~al.,}{{Aharonian}
  et~al.}{2006}]{2006A&A...460..743A}
{Aharonian} F.,  et~al., 2006, \mn@doi [\aap] {10.1051/0004-6361:20065940},
  \href {https://ui.adsabs.harvard.edu/abs/2006A&A...460..743A} {460, 743}

\bibitem[\protect\citeauthoryear{{Albert} et~al.,}{{Albert}
  et~al.}{2006}]{2006Sci...312.1771A}
{Albert} J.,  et~al., 2006, \mn@doi [Science] {10.1126/science.1128177}, \href
  {https://ui.adsabs.harvard.edu/abs/2006Sci...312.1771A} {312, 1771}

\bibitem[\protect\citeauthoryear{{Aliu} et~al.,}{{Aliu}
  et~al.}{2013}]{2013ApJ...779...88A}
{Aliu} E.,  et~al., 2013, \mn@doi [\apj] {10.1088/0004-637X/779/1/88}, \href
  {https://ui.adsabs.harvard.edu/abs/2013ApJ...779...88A} {779, 88}

\bibitem[\protect\citeauthoryear{{An} \& {Romani}}{{An} \&
  {Romani}}{2017}]{2017ApJ...838..145A}
{An} H.,  {Romani} R.~W.,  2017, \mn@doi [\apj] {10.3847/1538-4357/aa6623},
  \href {https://ui.adsabs.harvard.edu/abs/2017ApJ...838..145A} {838, 145}

\bibitem[\protect\citeauthoryear{{An} et~al.,}{{An}
  et~al.}{2015}]{2015ApJ...806..166A}
{An} H.,  et~al., 2015, \mn@doi [\apj] {10.1088/0004-637X/806/2/166}, \href
  {https://ui.adsabs.harvard.edu/abs/2015ApJ...806..166A} {806, 166}

\bibitem[\protect\citeauthoryear{{Ball} \& {Kirk}}{{Ball} \&
  {Kirk}}{2000}]{ball00}
{Ball} L.,  {Kirk} J.~G.,  2000, \mn@doi [Astroparticle Physics]
  {10.1016/S0927-6505(99)00112-7}, \href
  {https://ui.adsabs.harvard.edu/abs/2000APh....12..335B} {12, 335}

\bibitem[\protect\citeauthoryear{{Bednarek}}{{Bednarek}}{1997}]{1997A&A...322..523B}
{Bednarek} W.,  1997, \aap, \href
  {https://ui.adsabs.harvard.edu/abs/1997A&A...322..523B} {322, 523}

\bibitem[\protect\citeauthoryear{{Bogovalov}, {Khangulyan}, {Koldoba},
  {Ustyugova}  \& {Aharonian}}{{Bogovalov} et~al.}{2008}]{2008MNRAS.387...63B}
{Bogovalov} S.~V.,  {Khangulyan} D.~V.,  {Koldoba} A.~V.,  {Ustyugova} G.~V.,
  {Aharonian} F.~A.,  2008, \mn@doi [\mnras]
  {10.1111/j.1365-2966.2008.13226.x}, \href
  {https://ui.adsabs.harvard.edu/abs/2008MNRAS.387...63B} {387, 63}

\bibitem[\protect\citeauthoryear{{Bogovalov}, {Khangulyan}, {Koldoba},
  {Ustyugova}  \& {Aharonian}}{{Bogovalov} et~al.}{2019}]{2019MNRAS.490.3601B}
{Bogovalov} S.~V.,  {Khangulyan} D.,  {Koldoba} A.,  {Ustyugova} G.~V.,
  {Aharonian} F.,  2019, \mn@doi [\mnras] {10.1093/mnras/stz2815}, \href
  {https://ui.adsabs.harvard.edu/abs/2019MNRAS.490.3601B} {490, 3601}

\bibitem[\protect\citeauthoryear{{Bosch-Ramon}, {Barkov}, {Khangulyan}  \&
  {Perucho}}{{Bosch-Ramon} et~al.}{2012}]{2012A&A...544A..59B}
{Bosch-Ramon} V.,  {Barkov} M.~V.,  {Khangulyan} D.,   {Perucho} M.,  2012,
  \mn@doi [\aap] {10.1051/0004-6361/201219251}, \href
  {https://ui.adsabs.harvard.edu/abs/2012A&A...544A..59B} {544, A59}

\bibitem[\protect\citeauthoryear{{Canto}, {Raga}  \& {Wilkin}}{{Canto}
  et~al.}{1996}]{1996ApJ...469..729C}
{Canto} J.,  {Raga} A.~C.,   {Wilkin} F.~P.,  1996, \mn@doi [\apj]
  {10.1086/177820}, \href
  {https://ui.adsabs.harvard.edu/abs/1996ApJ...469..729C} {469, 729}

\bibitem[\protect\citeauthoryear{{Cerutti}, {Dubus}  \& {Henri}}{{Cerutti}
  et~al.}{2010}]{2010ASPC..422...41C}
{Cerutti} B.,  {Dubus} G.,   {Henri} G.,  2010, in {Mart{\'\i}} J.,
  {Luque-Escamilla} P.~L.,   {Combi} J.~A.,  eds,  Astronomical Society of the
  Pacific Conference Series Vol. 422, High Energy Phenomena in Massive Stars.
  p.~41

\bibitem[\protect\citeauthoryear{{Chang}, {Zhang}, {Ji}, {Chen}, {Kretschmar},
  {Kuulkers}, {Collmar}  \& {Liu}}{{Chang} et~al.}{2016}]{2016MNRAS.463..495C}
{Chang} Z.,  {Zhang} S.,  {Ji} L.,  {Chen} Y.~P.,  {Kretschmar} P.,  {Kuulkers}
  E.,  {Collmar} W.,   {Liu} C.~Z.,  2016, \mn@doi [\mnras]
  {10.1093/mnras/stw2009}, \href
  {https://ui.adsabs.harvard.edu/abs/2016MNRAS.463..495C} {463, 495}

\bibitem[\protect\citeauthoryear{{Chen}, {Takata}, {Yi}, {Yu}  \&
  {Cheng}}{{Chen} et~al.}{2019}]{2019A&A...627A..87C}
{Chen} A.~M.,  {Takata} J.,  {Yi} S.~X.,  {Yu} Y.~W.,   {Cheng} K.~S.,  2019,
  \mn@doi [\aap] {10.1051/0004-6361/201935166}, \href
  {https://ui.adsabs.harvard.edu/abs/2019A&A...627A..87C} {627, A87}

\bibitem[\protect\citeauthoryear{{Chernyakova}, {Neronov}, {Aharonian},
  {Uchiyama}  \& {Takahashi}}{{Chernyakova} et~al.}{2009}]{2009MNRAS.397.2123C}
{Chernyakova} M.,  {Neronov} A.,  {Aharonian} F.,  {Uchiyama} Y.,   {Takahashi}
  T.,  2009, \mn@doi [\mnras] {10.1111/j.1365-2966.2009.15116.x}, \href
  {https://ui.adsabs.harvard.edu/abs/2009MNRAS.397.2123C} {397, 2123}

\bibitem[\protect\citeauthoryear{{Chernyakova} et~al.,}{{Chernyakova}
  et~al.}{2015}]{2015MNRAS.454.1358C}
{Chernyakova} M.,  et~al., 2015, \mn@doi [\mnras] {10.1093/mnras/stv1988},
  \href {https://ui.adsabs.harvard.edu/abs/2015MNRAS.454.1358C} {454, 1358}

\bibitem[\protect\citeauthoryear{{Cominsky}}{{Cominsky}}{1994}]{1994surp.rept.....C}
{Cominsky} L.~R.,  1994, Technical report, {ROSAT observations of the binary
  Be-star and radio pulsar PSR1259-63}

\bibitem[\protect\citeauthoryear{{Corbet} et~al.,}{{Corbet}
  et~al.}{2016}]{2016ApJ...829..105C}
{Corbet} R.~H.~D.,  et~al., 2016, \mn@doi [\apj] {10.3847/0004-637X/829/2/105},
  \href {https://ui.adsabs.harvard.edu/abs/2016ApJ...829..105C} {829, 105}

\bibitem[\protect\citeauthoryear{{Corbet} et~al.,}{{Corbet}
  et~al.}{2019}]{corbet19}
{Corbet} R.~H.~D.,  et~al., 2019, arXiv e-prints, \href
  {https://ui.adsabs.harvard.edu/abs/2019arXiv190810764C} {p. arXiv:1908.10764}

\bibitem[\protect\citeauthoryear{{Davies}, {Elliott}  \& {Meaburn}}{{Davies}
  et~al.}{1976}]{1976MmRAS..81...89D}
{Davies} R.~D.,  {Elliott} K.~H.,   {Meaburn} J.,  1976, \memras, \href
  {https://ui.adsabs.harvard.edu/abs/1976MmRAS..81...89D} {81, 89}

\bibitem[\protect\citeauthoryear{{Dubus}}{{Dubus}}{2013}]{2013A&ARv..21...64D}
{Dubus} G.,  2013, \mn@doi [\aapr] {10.1007/s00159-013-0064-5}, \href
  {https://ui.adsabs.harvard.edu/abs/2013A&ARv..21...64D} {21, 64}

\bibitem[\protect\citeauthoryear{{Dubus}, {Cerutti}  \& {Henri}}{{Dubus}
  et~al.}{2010}]{2010A&A...516A..18D}
{Dubus} G.,  {Cerutti} B.,   {Henri} G.,  2010, \mn@doi [\aap]
  {10.1051/0004-6361/201014023}, \href
  {https://ui.adsabs.harvard.edu/abs/2010A&A...516A..18D} {516, A18}

\bibitem[\protect\citeauthoryear{{Gould} \& {Schr{\'e}der}}{{Gould} \&
  {Schr{\'e}der}}{1967}]{1967PhRv..155.1404G}
{Gould} R.~J.,  {Schr{\'e}der} G.~P.,  1967, \mn@doi [Physical Review]
  {10.1103/PhysRev.155.1404}, \href
  {https://ui.adsabs.harvard.edu/abs/1967PhRv..155.1404G} {155, 1404}

\bibitem[\protect\citeauthoryear{{H.~E.~S.~S. Collaboration}
  et~al.,}{{H.~E.~S.~S. Collaboration} et~al.}{2012}]{2012A&A...541A...5H}
{H.~E.~S.~S. Collaboration} et~al., 2012, \mn@doi [\aap]
  {10.1051/0004-6361/201218843}, \href
  {https://ui.adsabs.harvard.edu/abs/2012A&A...541A...5H} {541, A5}

\bibitem[\protect\citeauthoryear{{HESS Collaboration} et~al.,}{{HESS
  Collaboration} et~al.}{2018}]{2018A&A...610L..17H}
{HESS Collaboration} et~al., 2018, \mn@doi [\aap]
  {10.1051/0004-6361/201732426}, \href
  {https://ui.adsabs.harvard.edu/abs/2018A&A...610L..17H} {610, L17}

\bibitem[\protect\citeauthoryear{{Hibschman} \& {Arons}}{{Hibschman} \&
  {Arons}}{2001a}]{2001ApJ...554..624H}
{Hibschman} J.~A.,  {Arons} J.,  2001a, \mn@doi [\apj] {10.1086/321378}, \href
  {https://ui.adsabs.harvard.edu/abs/2001ApJ...554..624H} {554, 624}

\bibitem[\protect\citeauthoryear{{Hibschman} \& {Arons}}{{Hibschman} \&
  {Arons}}{2001b}]{2001ApJ...560..871H}
{Hibschman} J.~A.,  {Arons} J.,  2001b, \mn@doi [\apj] {10.1086/323069}, \href
  {https://ui.adsabs.harvard.edu/abs/2001ApJ...560..871H} {560, 871}

\bibitem[\protect\citeauthoryear{{Hinton} et~al.,}{{Hinton}
  et~al.}{2009}]{2009ApJ...690L.101H}
{Hinton} J.~A.,  et~al., 2009, \mn@doi [\apjl] {10.1088/0004-637X/690/2/L101},
  \href {https://ui.adsabs.harvard.edu/abs/2009ApJ...690L.101H} {690, L101}

\bibitem[\protect\citeauthoryear{{Ho}, {Ng}, {Lyne}, {Stappers}, {Coe},
  {Halpern}, {Johnson}  \& {Steele}}{{Ho} et~al.}{2017}]{2017MNRAS.464.1211H}
{Ho} W. C.~G.,  {Ng} C.~Y.,  {Lyne} A.~G.,  {Stappers} B.~W.,  {Coe} M.~J.,
  {Halpern} J.~P.,  {Johnson} T.~J.,   {Steele} I.~A.,  2017, \mn@doi [\mnras]
  {10.1093/mnras/stw2420}, \href
  {https://ui.adsabs.harvard.edu/abs/2017MNRAS.464.1211H} {464, 1211}

\bibitem[\protect\citeauthoryear{{Jauch} \& {Rohrlich}}{{Jauch} \&
  {Rohrlich}}{1976}]{1976tper.book.....J}
{Jauch} J.~M.,  {Rohrlich} F.,  1976, {The theory of photons and electrons. The
  relativistic quantum field theory of charged particles with spin one-half}

\bibitem[\protect\citeauthoryear{{Johnston}, {Manchester}, {Lyne}, {Bailes},
  {Kaspi}, {Qiao}  \& {D'Amico}}{{Johnston} et~al.}{1992}]{1992ApJ...387L..37J}
{Johnston} S.,  {Manchester} R.~N.,  {Lyne} A.~G.,  {Bailes} M.,  {Kaspi}
  V.~M.,  {Qiao} G.,   {D'Amico} N.,  1992, \mn@doi [\apjl] {10.1086/186300},
  \href {https://ui.adsabs.harvard.edu/abs/1992ApJ...387L..37J} {387, L37}

\bibitem[\protect\citeauthoryear{{Kapala}, {Bulik}, {Rudak}, {Dubus}  \&
  {Lyczek}}{{Kapala} et~al.}{2010}]{2010tsra.confE.193K}
{Kapala} M.,  {Bulik} T.,  {Rudak} B.,  {Dubus} G.,   {Lyczek} M.,  2010, in
  25th Texas Symposium on Relativistic Astrophysics. p.~193

\bibitem[\protect\citeauthoryear{{Kennel} \& {Coroniti}}{{Kennel} \&
  {Coroniti}}{1984a}]{1984ApJ...283..694K}
{Kennel} C.~F.,  {Coroniti} F.~V.,  1984a, \mn@doi [\apj] {10.1086/162356},
  \href {https://ui.adsabs.harvard.edu/abs/1984ApJ...283..694K} {283, 694}

\bibitem[\protect\citeauthoryear{{Kennel} \& {Coroniti}}{{Kennel} \&
  {Coroniti}}{1984b}]{1984ApJ...283..710K}
{Kennel} C.~F.,  {Coroniti} F.~V.,  1984b, \mn@doi [\apj] {10.1086/162357},
  \href {https://ui.adsabs.harvard.edu/abs/1984ApJ...283..710K} {283, 710}

\bibitem[\protect\citeauthoryear{{Khangulyan}, {Hnatic}  \&
  {Aharonian}}{{Khangulyan} et~al.}{2007}]{2007Ap&SS.309..261K}
{Khangulyan} D.,  {Hnatic} S.,   {Aharonian} F.,  2007, \mn@doi [\apss]
  {10.1007/s10509-007-9441-8}, \href
  {https://ui.adsabs.harvard.edu/abs/2007Ap&SS.309..261K} {309, 261}

\bibitem[\protect\citeauthoryear{{Khangulyan}, {Aharonian}  \&
  {Bosch-Ramon}}{{Khangulyan} et~al.}{2008}]{2008MNRAS.383..467K}
{Khangulyan} D.,  {Aharonian} F.,   {Bosch-Ramon} V.,  2008, \mn@doi [\mnras]
  {10.1111/j.1365-2966.2007.12572.x}, \href
  {https://ui.adsabs.harvard.edu/abs/2008MNRAS.383..467K} {383, 467}

\bibitem[\protect\citeauthoryear{{Komin}}{{Komin}}{2017}]{2017heas.confE..33K}
{Komin} N.,  2017, in 5th Annual Conference on High Energy Astrophysics in
  Southern Africa. p.~33

\bibitem[\protect\citeauthoryear{{Li}, {Torres}, {Cheng}, {de O{\~n}a
  Wilhelmi}, {Kretschmar}, {Hou}  \& {Takata}}{{Li}
  et~al.}{2017}]{2017ApJ...846..169L}
{Li} J.,  {Torres} D.~F.,  {Cheng} K.~S.,  {de O{\~n}a Wilhelmi} E.,
  {Kretschmar} P.,  {Hou} X.,   {Takata} J.,  2017, \mn@doi [\apj]
  {10.3847/1538-4357/aa7ff7}, \href
  {https://ui.adsabs.harvard.edu/abs/2017ApJ...846..169L} {846, 169}

\bibitem[\protect\citeauthoryear{{Long}, {Helfand}  \& {Grabelsky}}{{Long}
  et~al.}{1981}]{1981ApJ...248..925L}
{Long} K.~S.,  {Helfand} D.~J.,   {Grabelsky} D.~A.,  1981, \mn@doi [\apj]
  {10.1086/159222}, \href
  {https://ui.adsabs.harvard.edu/abs/1981ApJ...248..925L} {248, 925}

\bibitem[\protect\citeauthoryear{{Lyne}, {Stappers}, {Keith}, {Ray}, {Kerr},
  {Camilo}  \& {Johnson}}{{Lyne} et~al.}{2015}]{2015MNRAS.451..581L}
{Lyne} A.~G.,  {Stappers} B.~W.,  {Keith} M.~J.,  {Ray} P.~S.,  {Kerr} M.,
  {Camilo} F.,   {Johnson} T.~J.,  2015, \mn@doi [\mnras]
  {10.1093/mnras/stv236}, \href
  {https://ui.adsabs.harvard.edu/abs/2015MNRAS.451..581L} {451, 581}

\bibitem[\protect\citeauthoryear{{Macri}, {Stanek}, {Bersier}, {Greenhill}  \&
  {Reid}}{{Macri} et~al.}{2006}]{2006ApJ...652.1133M}
{Macri} L.~M.,  {Stanek} K.~Z.,  {Bersier} D.,  {Greenhill} L.~J.,   {Reid}
  M.~J.,  2006, \mn@doi [\apj] {10.1086/508530}, \href
  {https://ui.adsabs.harvard.edu/abs/2006ApJ...652.1133M} {652, 1133}

\bibitem[\protect\citeauthoryear{{Mariaud}, {Bordas}, {Aharonian}, {Dubus},
  {B{\"o}ttcher}, {de Naurois}, {Romoli}  \& {Zabalza}}{{Mariaud}
  et~al.}{2015}]{2015ICRC...34..885M}
{Mariaud} C.,  {Bordas} P.,  {Aharonian} F.,  {Dubus} G.,  {B{\"o}ttcher} M.,
  {de Naurois} M.,  {Romoli} C.,   {Zabalza} V.,  2015, in 34th International
  Cosmic Ray Conference (ICRC2015). p.~885

\bibitem[\protect\citeauthoryear{{Mart{\'\i}-Devesa} \&
  {Reimer}}{{Mart{\'\i}-Devesa} \& {Reimer}}{2020}]{2020arXiv200102701M}
{Mart{\'\i}-Devesa} G.,  {Reimer} O.,  2020, arXiv e-prints, \href
  {https://ui.adsabs.harvard.edu/abs/2020arXiv200102701M} {p. arXiv:2001.02701}

\bibitem[\protect\citeauthoryear{{Mold{\'o}n}, {Rib{\'o}}  \&
  {Paredes}}{{Mold{\'o}n} et~al.}{2012}]{2012A&A...548A.103M}
{Mold{\'o}n} J.,  {Rib{\'o}} M.,   {Paredes} J.~M.,  2012, \mn@doi [\aap]
  {10.1051/0004-6361/201219553}, \href
  {https://ui.adsabs.harvard.edu/abs/2012A&A...548A.103M} {548, A103}

\bibitem[\protect\citeauthoryear{{Moritani}, {Kawano}, {Chimasu}, {Kawachi},
  {Takahashi}, {Takata}  \& {Carciofi}}{{Moritani}
  et~al.}{2018}]{2018PASJ...70...61M}
{Moritani} Y.,  {Kawano} T.,  {Chimasu} S.,  {Kawachi} A.,  {Takahashi} H.,
  {Takata} J.,   {Carciofi} A.~C.,  2018, \mn@doi [\pasj]
  {10.1093/pasj/psy053}, \href
  {https://ui.adsabs.harvard.edu/abs/2018PASJ...70...61M} {70, 61}

\bibitem[\protect\citeauthoryear{{Pietrzy{\'n}ski} et~al.,}{{Pietrzy{\'n}ski}
  et~al.}{2013}]{2013EAS....64..305P}
{Pietrzy{\'n}ski} G.,  et~al., 2013, in {Pavlovski} K.,  {Tkachenko} A.,
  {Torres} G.,  eds,  Vol. 64, EAS Publications Series. pp 305--307,
  \mn@doi{10.1051/eas/1364042}

\bibitem[\protect\citeauthoryear{{Puls} et~al.,}{{Puls}
  et~al.}{1996}]{1996A&A...305..171P}
{Puls} J.,  et~al., 1996, \aap, \href
  {https://ui.adsabs.harvard.edu/abs/1996A&A...305..171P} {305, 171}

\bibitem[\protect\citeauthoryear{{Seward}, {Charles}, {Foster}, {Dickel},
  {Romero}, {Edwards}, {Perry}  \& {Williams}}{{Seward}
  et~al.}{2012}]{2012ApJ...759..123S}
{Seward} F.~D.,  {Charles} P.~A.,  {Foster} D.~L.,  {Dickel} J.~R.,  {Romero}
  P.~S.,  {Edwards} Z.~I.,  {Perry} M.,   {Williams} R.~M.,  2012, \mn@doi
  [\apj] {10.1088/0004-637X/759/2/123}, \href
  {https://ui.adsabs.harvard.edu/abs/2012ApJ...759..123S} {759, 123}

\bibitem[\protect\citeauthoryear{{Sierpowska-Bartosik} \&
  {Bednarek}}{{Sierpowska-Bartosik} \& {Bednarek}}{2008}]{2008MNRAS.385.2279S}
{Sierpowska-Bartosik} A.,  {Bednarek} W.,  2008, \mn@doi [\mnras]
  {10.1111/j.1365-2966.2008.13002.x}, \href
  {https://ui.adsabs.harvard.edu/abs/2008MNRAS.385.2279S} {385, 2279}

\bibitem[\protect\citeauthoryear{{Sierpowska-Bartosik} \&
  {Torres}}{{Sierpowska-Bartosik} \& {Torres}}{2007}]{2007ApJ...671L.145S}
{Sierpowska-Bartosik} A.,  {Torres} D.~F.,  2007, \mn@doi [\apjl]
  {10.1086/525041}, \href
  {https://ui.adsabs.harvard.edu/abs/2007ApJ...671L.145S} {671, L145}

\bibitem[\protect\citeauthoryear{{Sierpowska-Bartosik} \&
  {Torres}}{{Sierpowska-Bartosik} \& {Torres}}{2008}]{2008AIPC.1085..253S}
{Sierpowska-Bartosik} A.,  {Torres} D.~F.,  2008, in {Aharonian} F.~A.,
  {Hofmann} W.,   {Rieger} F.,  eds,  Vol. 1085, American Institute of Physics
  Conference Series. pp 253--256, \mn@doi{10.1063/1.3076653}

\bibitem[\protect\citeauthoryear{{Takahashi} et~al.,}{{Takahashi}
  et~al.}{2009}]{2009ApJ...697..592T}
{Takahashi} T.,  et~al., 2009, \mn@doi [\apj] {10.1088/0004-637X/697/1/592},
  \href {https://ui.adsabs.harvard.edu/abs/2009ApJ...697..592T} {697, 592}

\bibitem[\protect\citeauthoryear{{Takata} \& {Taam}}{{Takata} \&
  {Taam}}{2009}]{2009ApJ...702..100T}
{Takata} J.,  {Taam} R.~E.,  2009, \mn@doi [\apj]
  {10.1088/0004-637X/702/1/100}, \href
  {https://ui.adsabs.harvard.edu/abs/2009ApJ...702..100T} {702, 100}

\bibitem[\protect\citeauthoryear{{Takata}, {Leung}, {Tam}, {Kong}, {Hui}  \&
  {Cheng}}{{Takata} et~al.}{2014}]{2014ApJ...790...18T}
{Takata} J.,  {Leung} G. C.~K.,  {Tam} P.~H.~T.,  {Kong} A.~K.~H.,  {Hui}
  C.~Y.,   {Cheng} K.~S.,  2014, \mn@doi [\apj] {10.1088/0004-637X/790/1/18},
  \href {https://ui.adsabs.harvard.edu/abs/2014ApJ...790...18T} {790, 18}

\bibitem[\protect\citeauthoryear{{Takata}, {Tam}, {Ng}, {Li}, {Kong}, {Hui}  \&
  {Cheng}}{{Takata} et~al.}{2017}]{2017ApJ...836..241T}
{Takata} J.,  {Tam} P.~H.~T.,  {Ng} C.~W.,  {Li} K.~L.,  {Kong} A.~K.~H.,
  {Hui} C.~Y.,   {Cheng} K.~S.,  2017, \mn@doi [\apj]
  {10.3847/1538-4357/aa5c80}, \href
  {https://ui.adsabs.harvard.edu/abs/2017ApJ...836..241T} {836, 241}

\bibitem[\protect\citeauthoryear{{Tam}, {Huang}, {Takata}, {Hui}, {Kong}  \&
  {Cheng}}{{Tam} et~al.}{2011}]{tam11}
{Tam} P.~H.~T.,  {Huang} R.~H.~H.,  {Takata} J.,  {Hui} C.~Y.,  {Kong}
  A.~K.~H.,   {Cheng} K.~S.,  2011, \mn@doi [\apjl]
  {10.1088/2041-8205/736/1/L10}, \href
  {https://ui.adsabs.harvard.edu/abs/2011ApJ...736L..10T} {736, L10}

\bibitem[\protect\citeauthoryear{{Tam}, {He}, {Pal}  \& {Cui}}{{Tam}
  et~al.}{2018}]{tam18}
{Tam} P.~H.~T.,  {He} X.~B.,  {Pal} P.~S.,   {Cui} Y.,  2018, \mn@doi [\apj]
  {10.3847/1538-4357/aacf00}, \href
  {https://ui.adsabs.harvard.edu/abs/2018ApJ...862..165T} {862, 165}

\bibitem[\protect\citeauthoryear{{Tam} et~al.,}{{Tam}
  et~al.}{2020}]{2020arXiv200107138T}
{Tam} P.-H.~T.,  et~al., 2020, arXiv e-prints, \href
  {https://ui.adsabs.harvard.edu/abs/2020arXiv200107138T} {p. arXiv:2001.07138}

\bibitem[\protect\citeauthoryear{{Tang}}{{Tang}}{2018}]{2018Ap&SS.363...25T}
{Tang} Q.-W.,  2018, \mn@doi [\apss] {10.1007/s10509-017-3243-4}, \href
  {https://ui.adsabs.harvard.edu/abs/2018Ap%26SS.363...25T} {363, 25}

\bibitem[\protect\citeauthoryear{{Tang}, {Peng}, {Liu}, {Tam}  \&
  {Wang}}{{Tang} et~al.}{2017}]{2017ApJ...843...42T}
{Tang} Q.-W.,  {Peng} F.-K.,  {Liu} R.-Y.,  {Tam} P.-H.~T.,   {Wang} X.-Y.,
  2017, \mn@doi [\apj] {10.3847/1538-4357/aa7464}, \href
  {https://ui.adsabs.harvard.edu/abs/2017ApJ...843...42T} {843, 42}

\bibitem[\protect\citeauthoryear{{Tavani} \& {Arons}}{{Tavani} \&
  {Arons}}{1997}]{1997ApJ...477..439T}
{Tavani} M.,  {Arons} J.,  1997, \mn@doi [\apj] {10.1086/303676}, \href
  {https://ui.adsabs.harvard.edu/abs/1997ApJ...477..439T} {477, 439}

\bibitem[\protect\citeauthoryear{{Timokhin} \& {Harding}}{{Timokhin} \&
  {Harding}}{2015}]{2015ApJ...810..144T}
{Timokhin} A.~N.,  {Harding} A.~K.,  2015, \mn@doi [\apj]
  {10.1088/0004-637X/810/2/144}, \href
  {https://ui.adsabs.harvard.edu/abs/2015ApJ...810..144T} {810, 144}

\bibitem[\protect\citeauthoryear{{Timokhin} \& {Harding}}{{Timokhin} \&
  {Harding}}{2019}]{2019ApJ...871...12T}
{Timokhin} A.~N.,  {Harding} A.~K.,  2019, \mn@doi [\apj]
  {10.3847/1538-4357/aaf050}, \href
  {https://ui.adsabs.harvard.edu/abs/2019ApJ...871...12T} {871, 12}

\bibitem[\protect\citeauthoryear{{Torres}}{{Torres}}{2011}]{2011ASSP...21..531T}
{Torres} D.~F.,  2011, Astrophysics and Space Science Proceedings, \href
  {https://ui.adsabs.harvard.edu/abs/2011ASSP...21..531T} {21, 531}

\bibitem[\protect\citeauthoryear{{Walder}, {Folini}  \& {Meynet}}{{Walder}
  et~al.}{2012}]{2012SSRv..166..145W}
{Walder} R.,  {Folini} D.,   {Meynet} G.,  2012, \mn@doi [\ssr]
  {10.1007/s11214-011-9771-2}, \href
  {https://ui.adsabs.harvard.edu/abs/2012SSRv..166..145W} {166, 145}

\bibitem[\protect\citeauthoryear{{Wood}, {Caputo}, {Charles}, {Di Mauro},
  {Magill}, {Perkins}  \& {Fermi-LAT Collaboration}}{{Wood}
  et~al.}{2017}]{2017ICRC...35..824W}
{Wood} M.,  {Caputo} R.,  {Charles} E.,  {Di Mauro} M.,  {Magill} J.,
  {Perkins} J.~S.,   {Fermi-LAT Collaboration} 2017, in 35th International
  Cosmic Ray Conference (ICRC2017). p.~824 (\mn@eprint {arXiv} {1707.09551})

\bibitem[\protect\citeauthoryear{{Yamaguchi} \& {Takahara}}{{Yamaguchi} \&
  {Takahara}}{2012}]{2012ApJ...761..146Y}
{Yamaguchi} M.~S.,  {Takahara} F.,  2012, \mn@doi [\apj]
  {10.1088/0004-637X/761/2/146}, \href
  {https://ui.adsabs.harvard.edu/abs/2012ApJ...761..146Y} {761, 146}

\bibitem[\protect\citeauthoryear{{Zabalza}, {Bosch-Ramon}, {Aharonian}  \&
  {Khangulyan}}{{Zabalza} et~al.}{2013}]{2013A&A...551A..17Z}
{Zabalza} V.,  {Bosch-Ramon} V.,  {Aharonian} F.,   {Khangulyan} D.,  2013,
  \mn@doi [\aap] {10.1051/0004-6361/201220589}, \href
  {https://ui.adsabs.harvard.edu/abs/2013A&A...551A..17Z} {551, A17}

\bibitem[\protect\citeauthoryear{{de Grijs}, {Wicker}  \& {Bono}}{{de Grijs}
  et~al.}{2014}]{2014AJ....147..122D}
{de Grijs} R.,  {Wicker} J.~E.,   {Bono} G.,  2014, \mn@doi [\aj]
  {10.1088/0004-6256/147/5/122}, \href
  {https://ui.adsabs.harvard.edu/abs/2014AJ....147..122D} {147, 122}

\bibitem[\protect\citeauthoryear{{van Soelen}, {Komin}, {Kniazev}  \&
  {V{\"a}is{\"a}nen}}{{van Soelen} et~al.}{2019}]{2019MNRAS.484.4347V}
{van Soelen} B.,  {Komin} N.,  {Kniazev} A.,   {V{\"a}is{\"a}nen} P.,  2019,
  \mn@doi [\mnras] {10.1093/mnras/stz289}, \href
  {https://ui.adsabs.harvard.edu/abs/2019MNRAS.484.4347V} {484, 4347}

\makeatother
\end{thebibliography}
\label{lastpage}
\end{document}